\documentclass{pasj00}
\draft

\begin{document}
\SetRunningHead{M. Makita \& S. Mineshige}{Emission-Line Eclipse Mapping
of Dwarf-Nova Accretion Disk}
\Received{2002/02/21}%{yyyy/mm/dd}
\Accepted{2002/04/02}%{yyyy/mm/dd}

\title{Emission-Line Eclipse Mapping of Velocity Fields 
in a Dwarf-Nova Accretion Disk}

\author{Makoto \textsc{Makita}}
\affil{Department of Astronomy, Kyoto University, Sakyo-ku, Kyoto
606-8502}
\email{makoto@kusastro.kyoto-u.ac.jp}

\and
\author{Shin \textsc{Mineshige}}
\affil{Yukawa Institute for Theoretical Physics, Kyoto University, Sakyo-ku, Kyoto
606-8502}
\email{minesige@yukawa.kyoto-u.ac.jp}

 %Do NOT move this preamble from here!

\maketitle

\begin{abstract}
We propose a new method, {\it emission-line eclipse mapping},
 to map the velocity fields of an accretion disk in position space.
Quiescent dwarf novae usually exhibit double-peaked emission line
profiles because of disk rotation.  Since
a part of a disk having a different line-of-sight velocity is 
successively obscured by a companion in eclipsing systems,
they show time-varying line profiles.
We calculated the time changes of the emission-line profiles, 
assuming Keplerian rotation fields ($v_\varphi\propto r^{-1/2}$
with $r$ being the distance from the disk center)
and an emissivity distribution of $j \propto r^{-3/2}$.
We, then, applied the usual eclipse mapping technique to the light curves
at each of 12--24 wavelengths across the line center to map
the region with the same line-of-sight velocity.
The reconstructed images typically exhibit a `two-eye' pattern 
for high line-of-sight velocities, and we can recover the relation, 
$v_\varphi \propto d^{-1/2}$, on the assumption of an axisymmetric disk,
where $d$ is the separation between the two {\lq}eyes{\rq}.
We will be able to probe the Keplerian rotation law,
the most fundamental assumption adopted in many disk models,
by high-speed spectroscopic observations with 8-m class telescopes.
\end{abstract}

\section{Introduction}
It is widely recognized that accretion disks play principal roles 
in various objects, 
including protostars, cataclysmic variables (CVs), X-ray binaries (XBs), 
and active galactic nuclei.
Among them, CVs are the most ideal objects for accretion-disk research
for the following reasons.  They are composed of a mass-losing, 
Roche-lobe-filling late-type star and an accreting white dwarf,
and their orbital periods typically range
 between 1.5 and 10 hr.  
Such timescales are very convenient for ground-based observations.
They are numerous; in other words, there are many CVs in our neighborhood.
Also, CVs are bright in optical bands and 
exhibit a variety of time variations, some of which 
are closely related to various accretion-disk processes.
Unlike XBs, moreover, the irradiation flux from the central parts of the disk
and from the central objects is not very large, which makes
accretion physics much simpler than otherwise.
Hence, CVs are useful laboratories to investigate accretion-disk
 physics.
However, the size of the accretion disks in CVs is typically 
$\sim R_\odot$, 
which is too small to resolve with ordinary telescopes; that is,
we always collect all of the photons emitted from the entire disk surface
(except for some special cases of young stellar objects).
To test the disk theory with observations, therefore,
we are obliged to compare the results only in terms of the integrated values,
which cannot uniquely constrain the spatial structure of accretion disks.

There exists, however, a novel method to indirectly `resolve' the surface
brightness distribution over the disk plane.
That is eclipse mapping (Horne \yearcite{H85}, \yearcite{H93}), 
which makes use of the eclipse of high-inclination binary systems.
Since a part of the disk is successively obscured by the secondary star 
during an eclipse, the eclipsing light curves contain information 
regarding the brightness distribution over the disk plane.  
We can reconstruct a two-dimensional (2D) brightness distribution
from one-dimensional eclipse light curves based on the concept
of the maximum entropy.
[As for the case of quasars, one may
use microlens events by stars in intervening galaxies 
(e.g., Yonehara et al. \yearcite{YMMFUT98}), 
but this method is irrelevant here.]
Similarly important is the technique of Doppler tomography 
(Marsh, Horne \yearcite{MH88}). 
Since the line profiles reflect velocity fields within the binary systems,
including those of the secondary star and the gas stream to the disk,
we can map the line-emissivity distribution in two-dimensional velocity space
by analyzing the line-profile variations during the course of the orbital motions.
This method, however, provides no definite information regarding
the velocity fields in position (real) space.

We, here, consider a third method, by combining the above two methods,
to map {\it the velocity fields in position space over the disk plane.}
The goal is to verify whether the disk rotation velocity ($v_\varphi$) 
simply obeys the Kepler's law, $v_{\rm K} = \sqrt{GM_1/r}$
(with $M_1$ and $r$ being the mass of a compact object and
the distance from the central object, respectively).
Moreover, we may be able to detect deviations from Keplerian rotation
due to spiral shock patterns (Sawada et al. 
\yearcite{SMH86a}a, b)
and/or non-negligible magnetic fields.
Thus, we can say that mapping the disk velocity fields in position space
is one of the most fundamental tests of the basic disk models.
Nevertheless, nobody has yet completed such a method,
to our best knowledge, partly because for that purpose we need to detect
changes of emission-line profiles during eclipse (lasting only several
tens of minutes), which requires an enormous amount of photons,  
too much to collect with 4-m class telescopes.
Recently built 8-m class telescopes can do this job, however.
In this paper, we describe the basic methodology and a simple test to probe
the disk velocity fields by using the eclipse of emission-line profiles
during the quiescence of dwarf novae.

Note that an apparently similar method has been proposed by 
Vrielmann et al. (\yearcite{V95});
however, no detailed description is available and
here we present an independent, slightly different method of our own.
In section 2, we give basic considerations
and then present the adopted model assumptions and procedures.
Next, we explain our calculations of model light curves, assuming Keplerian
rotation velocity fields, and present the resultant one-dimensional
velocity profile obtained by the proposed method in section 3.
The final section is devoted to discussion.

\section{Model and Procedure}
\subsection{Basic Considerations}
Dwarf novae (DNe) are a subclass of CVs
and are unique in  showing repetitive outbursts.
During the quiescent states between outbursts, DNe usually exhibit
strong, double-peaked emission lines
due to Doppler effects caused by disk rotation.  Therefore,
each line profile contains information about the disk velocity fields,
although we cannot spatially resolve them.
In eclipsing systems, however,
a part of the disk, which has a different line-of-sight velocity,
is successively obscured by a companion star,
giving rise to time variations of the emission line profiles.
We then apply the usual eclipse mapping technique to the radiation flux
at each of multiple wavelengths across the line center to obtain
the emissivity distribution of the region moving
with the corresponding line-of-sight velocity on the 2D disk plane.
With the radial emissivity distribution being specified, we are able to
obtain the rotational velocity as a function of radius
on the assumption of circular motion.

\begin{figure}
  \begin{center}
    \FigureFile(75mm,75mm){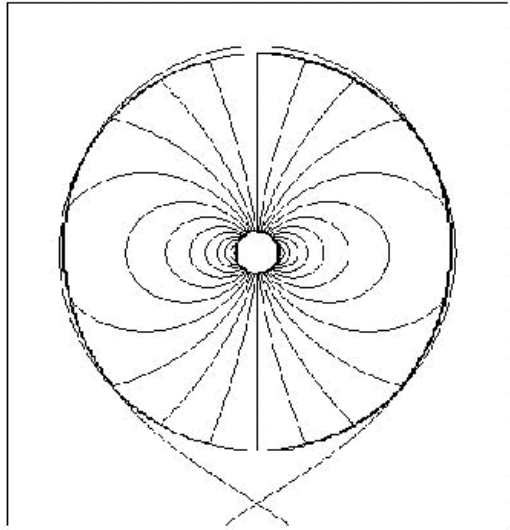}
  \end{center}
\caption{Contours of the constant line-of-sight velocity 
         on the disk plane for the case of an inclination of
 $i=80^\circ$ and a mass ratio of $q=0.5$. This picture corresponds to the
 case of $N=12$, where $N$ is the number of wavelength bins.}
\label{fig:kepconst}
\end{figure}

According to Horne and Marsh (\yearcite{HM86})
and Marsh and Horne (\yearcite{MH88}), who calculated and displayed
the contours of the constant line-of-sight velocity 
($v_{\rm l.o.s.}$) on the disk plane and the corresponding wavelengths
($\delta\lambda/\lambda_0 = v_{\rm l.o.s.}/c$ with 
$\lambda_0$ being the line center wavelength),
the shapes of the contours are distinct for
high and low line-of-sight velocities:
The large-$v_{\rm l.o.s.}$ contours exhibit `two-eye' patterns and
their separation increases with a decreasing line-of-sight velocity.
The small-$v_{\rm l.o.s.}$ contours, on the other hand, show
more complex patterns similar to the shape of a dipole field.
For a reference, we display the contours of the constant line-of-sight
velocity in Keplerian rotation viewed at the binary phase 0.0 for the case 
of an inclination angle of $i=80^\circ$ in figure \ref{fig:kepconst}.
The critical velocity discriminating these two different patterns is given by
$\pm v_\varphi (R_{\rm out}) \sin i$, 
with $R_{\rm out}$ being the radius of the outer edge of the disk,
which corresponds to the wavelengths 
of the two (blue and red) peaks in the line profile.
Since the large-$v_{\rm l.o.s.}$ parts are much easier to deal with,
we, here, aim to map the emissivity distribution of those parts.
In other words, we focus on
an analysis of the wing parts of the double-horned line profiles.
Then, as mentioned above, the relation between the line-of-sight velocity
and the separation between the two `eyes' (referred to as $d$)
provides information regarding
the disk rotational velocity fields on the axisymmetric assumption.
Since $i$, $M_1$, and other binary parameters are 
not usually precisely determined, we do not pay much attention
to the absolute values of $v_{\rm l.o.s.}$ but are concerned with
the relative relationship (e.g., a power-law index, 
if represented as a power-law) between $v_{\rm l.o.s.}$ and $d$.

\subsection{Disk Model}
We calculated the line-profile variations during an eclipse for a simple case
and tested our method to see to what extent 
we can derive information regarding velocity fields.
We considered a semi-detached binary system composed of a primary star 
with a mass of $M_1$ and a secondary star with a mass of $M_2$.
We assumed: 
(1) a geometrically thin accretion disk is formed around the primary,
(2) disk materials are orbiting on the circular orbits around the primary
with the Keplerian velocity,
and 
(3) the secondary star fills its Roche lobe. 
In calculations, we set
the binary separation, $a$, to be a unit of the length scale.
We further assumed a line emissivity of
\begin{equation}
 j(r)=\left(\frac{r}{R_{\rm in}}\right)^{-\frac{3}{2}}
\end{equation}
(see Horne, Saar \yearcite{HS91}; Horne \yearcite{H94})
between the radius of the inner edge of the disk, $R_{\rm in}=0.05$, 
and that of the outer edge, $R_{\rm out}=0.5$ in units of
binary separation, $a$.
(If we perform Doppler tomography at the same time, 
it is not necessary to assume $j(r)$, since
we can derive it on the assumption of the axisymmetric disk.)

As for the binary system parameters, we adopted a mass ratio of
$q \equiv M_2/M_1 = 0.5$, a binary separation of $a=R_\odot$,
and an inclination angle of $i=80^{\circ}$.
In physical units, the Keplerian rotation velocity is
\begin{equation}
  v_{\rm K} = \sqrt{\frac{GM_1}{r}}\sim 
  1000 \left(\frac{M_1}{M_\odot}\right)^{1/2}
       \left(\frac{r}{10^{10}{\rm cm}}\right)^{-1/2}~{\rm km~s}^{-1}. 
\end{equation}
That is, we needed to cover velocity ranges of
$v \simeq$ (1000--3000) $\sin i~ $km~s$^{-1}$
for the radius range of $r \simeq 10^9$--$10^{10}$cm and $M_1\sim M_\odot$.
Corresponding Doppler shifts are
$\delta\lambda/\lambda_0 = v \sin i/c \simeq (0.003$--$0.01) \sin i$ 
(with $c$ being the speed of light).

\subsection{Procedure of the Emission-Line Eclipse Mapping}
We divided the emission line profile into $N=$12 or 24 sections 
with a constant wavelength interval, $\Delta \lambda/\lambda_0 = 0.014/N$.
We calculated the continuum flux with an equivalent width (EW) of 100\AA, and
each line flux was normalized with this continuum flux.
In the following, $F_{\rm cont} = 1.0$ and $F_{\rm line}(\lambda)$ refers to 
the normalized line flux with respect to $F_{\rm cont}$.

For each section, an eclipse light curve was constructed from the binary
phase $\phi=-0.15$ to $\phi=+0.15$ with 51 phase bins.
We summed up the emission of the two sections (i.e. on the blue and red sides),
which have the same absolute line-of-sight velocity, to
increase the number of photons
and normalized it with its maximum value.
In some model, we only used either side of the emission profile,
since that would be better when the line profile is highly asymmetric.
Then, we made an eclipse map 
using the PRIDA code (Baptista, Steiner \yearcite{BS91}, \yearcite{BS93}) 
for each of $N$ calculated light curves with different wavelengths.
As for the default map, we, here, simply used an axisymmetric map
(discussed later).
We fixed the inclination angle to be $i=80^\circ$.

\section{Reconstructed Images and Velocity Fields}
\subsection{Basic Model}

\begin{figure*}
 \begin{center}
\FigureFile(40mm,35mm){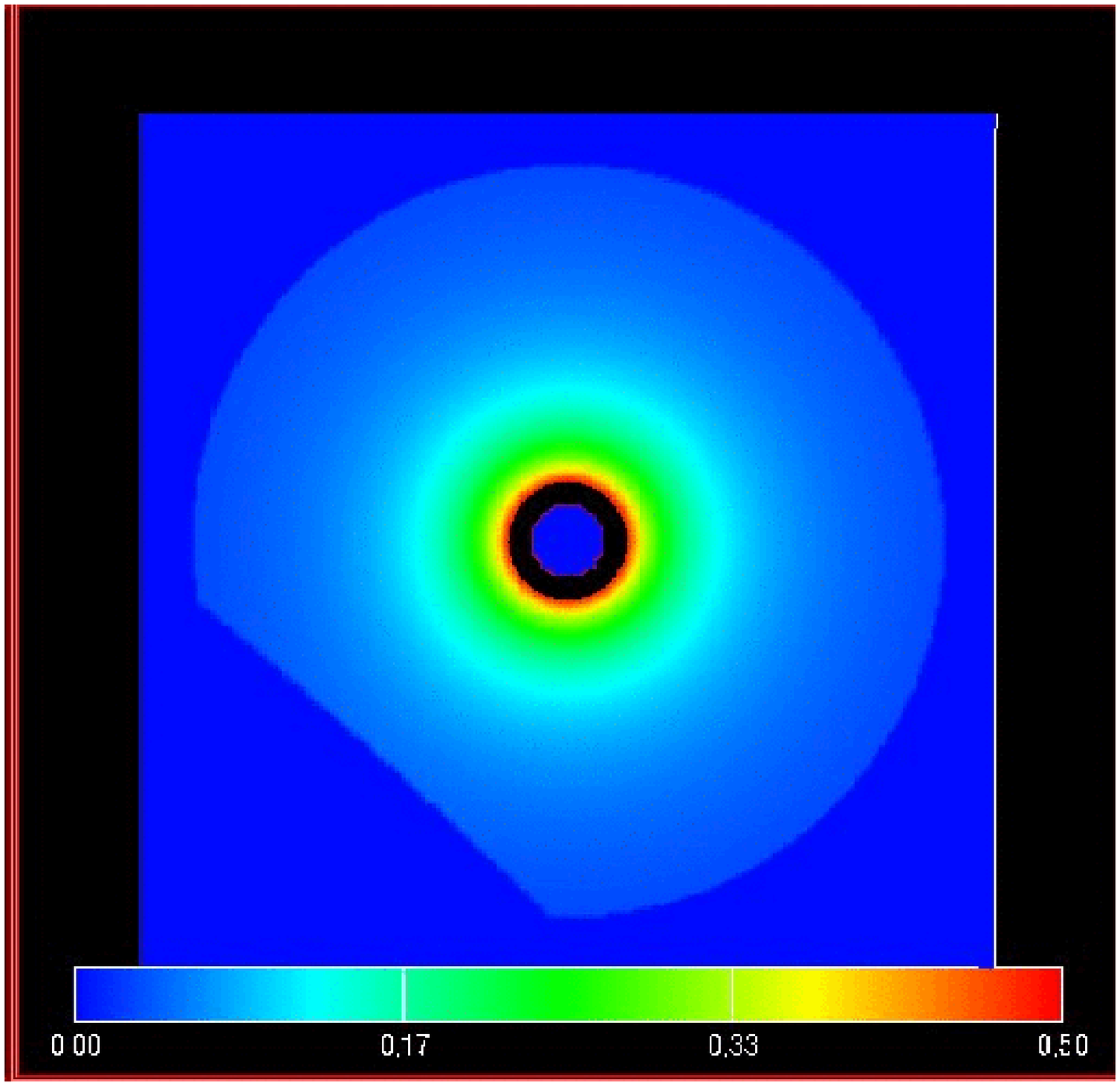}
\FigureFile(40mm,45mm){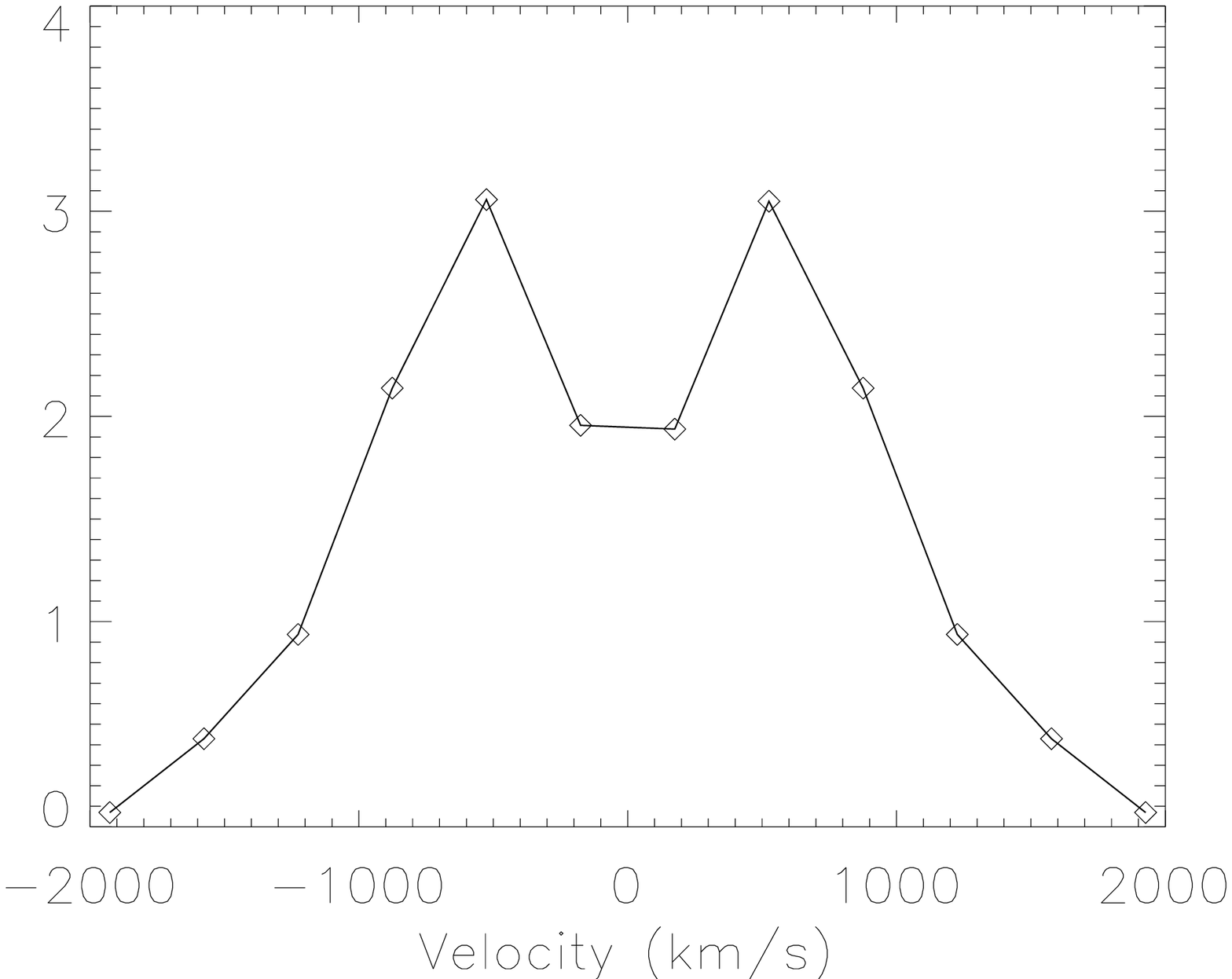}
\FigureFile(40mm,45mm){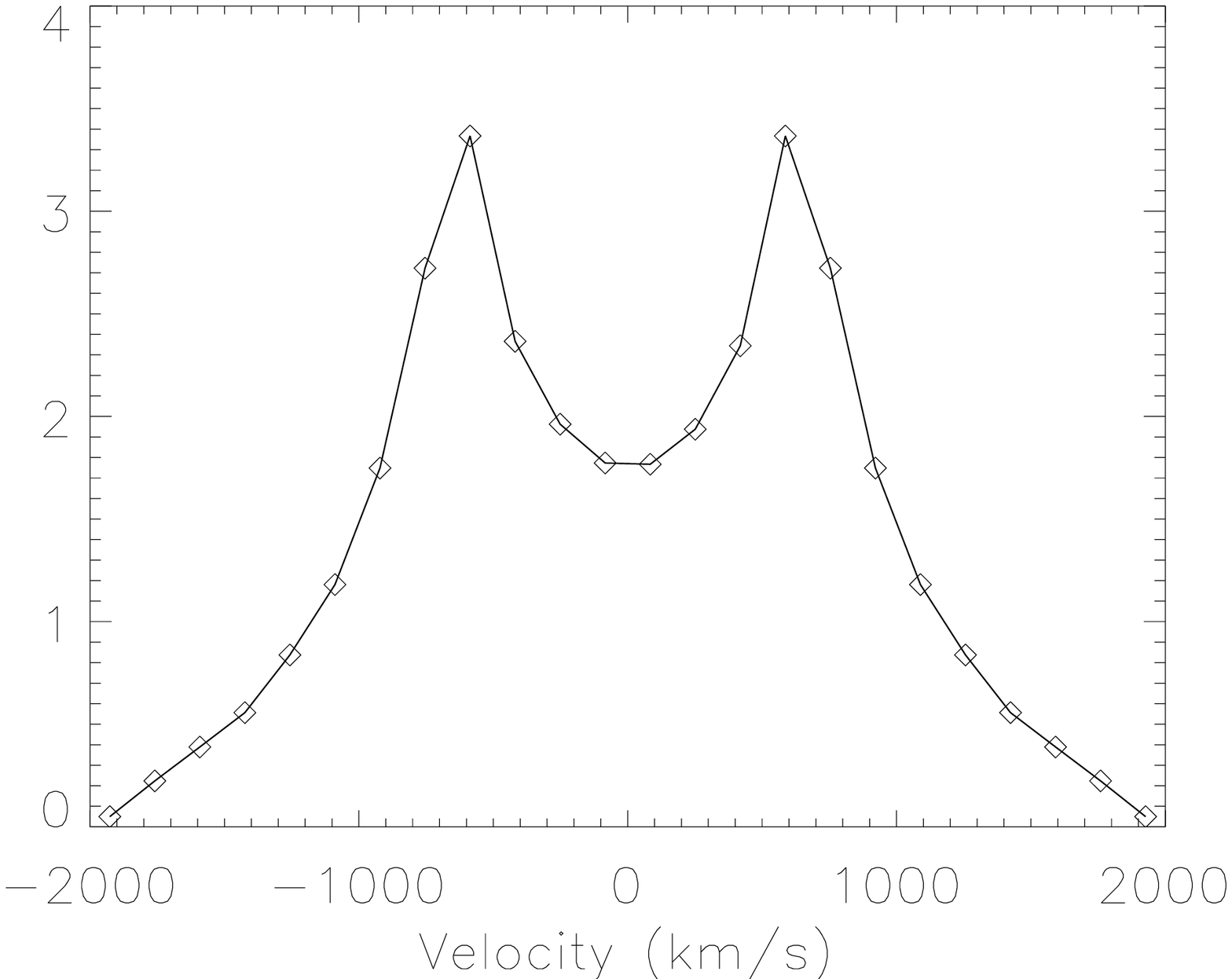}\\
\FigureFile(40mm,35mm){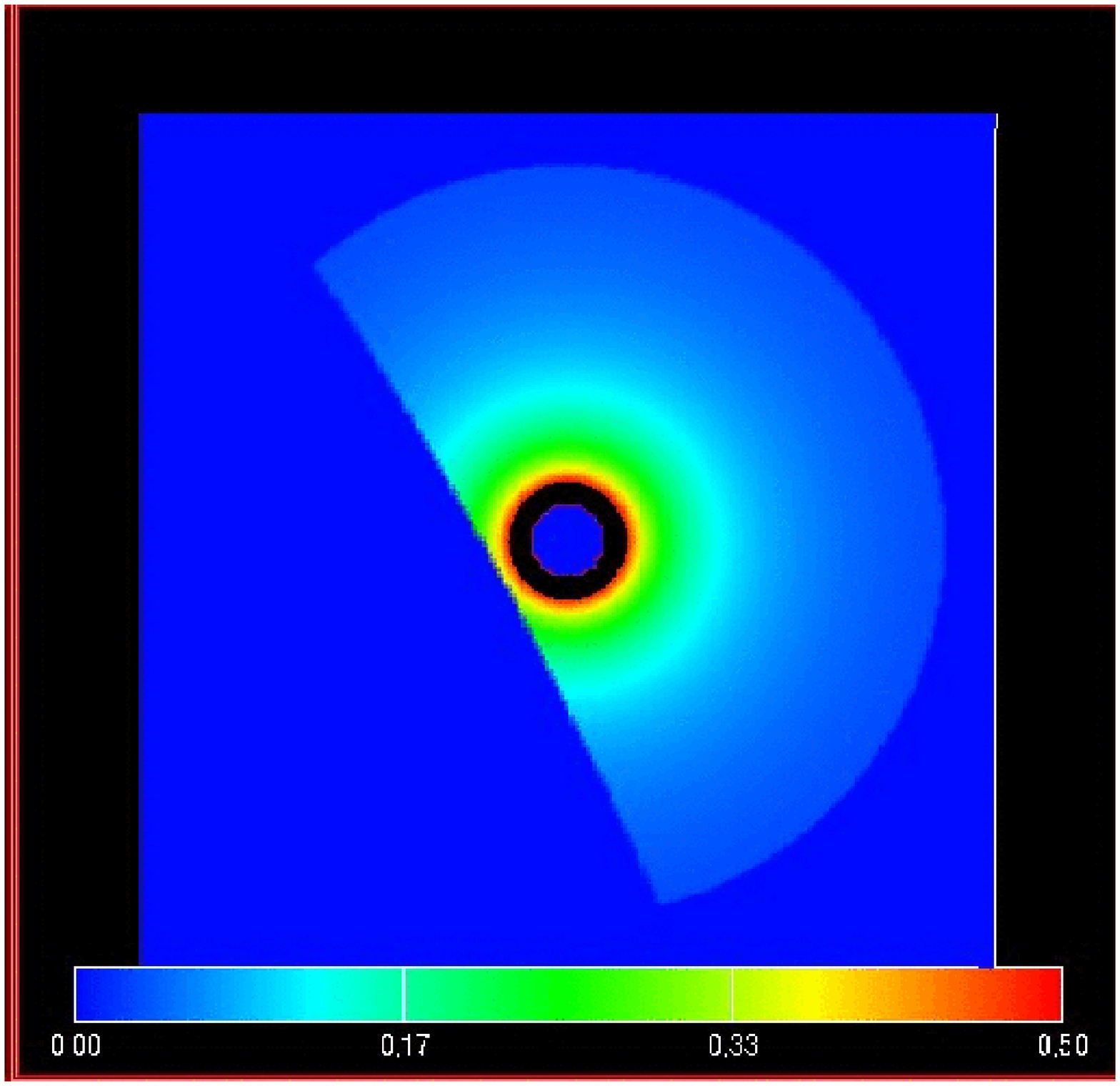}
\FigureFile(40mm,45mm){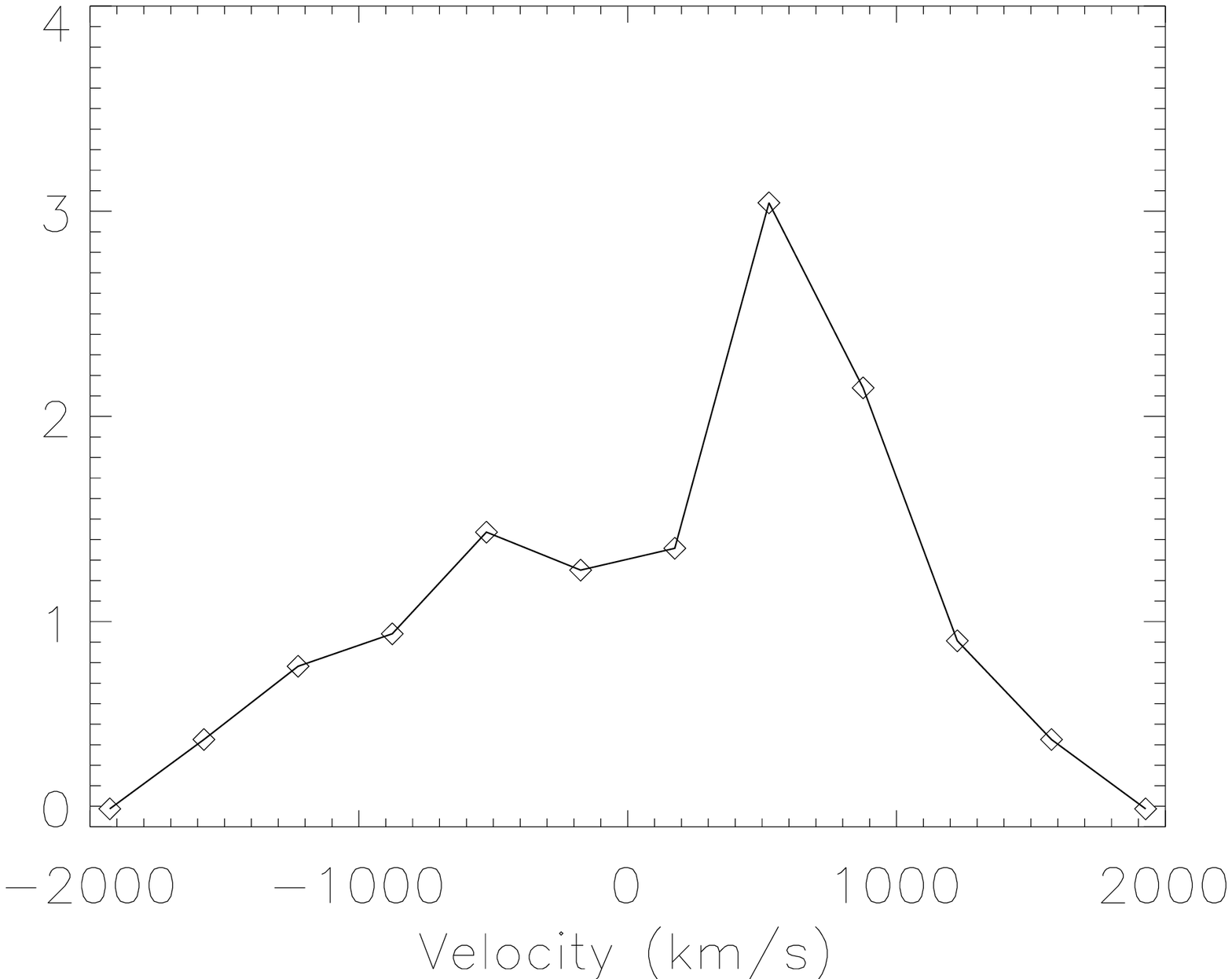}
\FigureFile(40mm,45mm){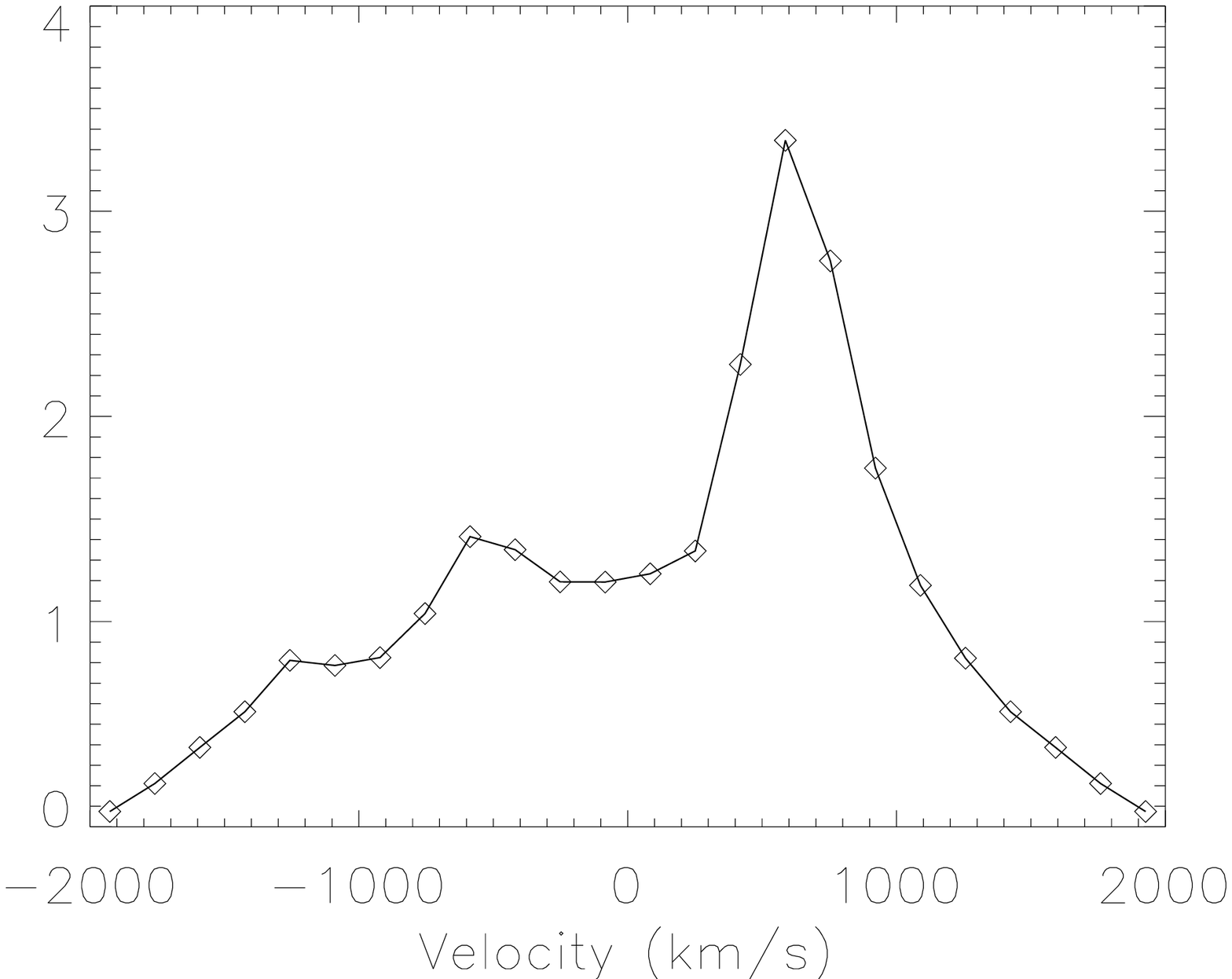}\\
\FigureFile(40mm,35mm){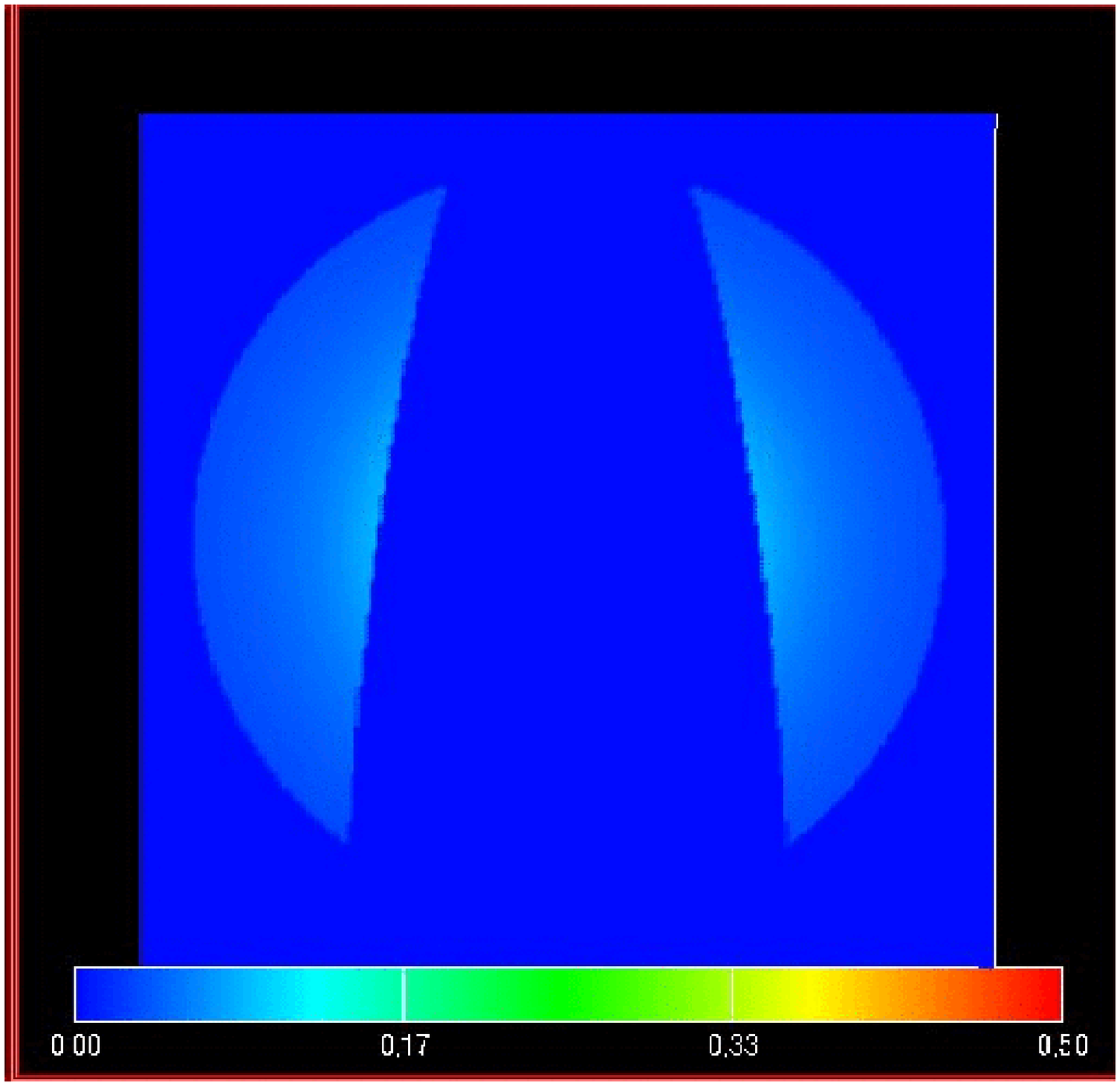}
\FigureFile(40mm,45mm){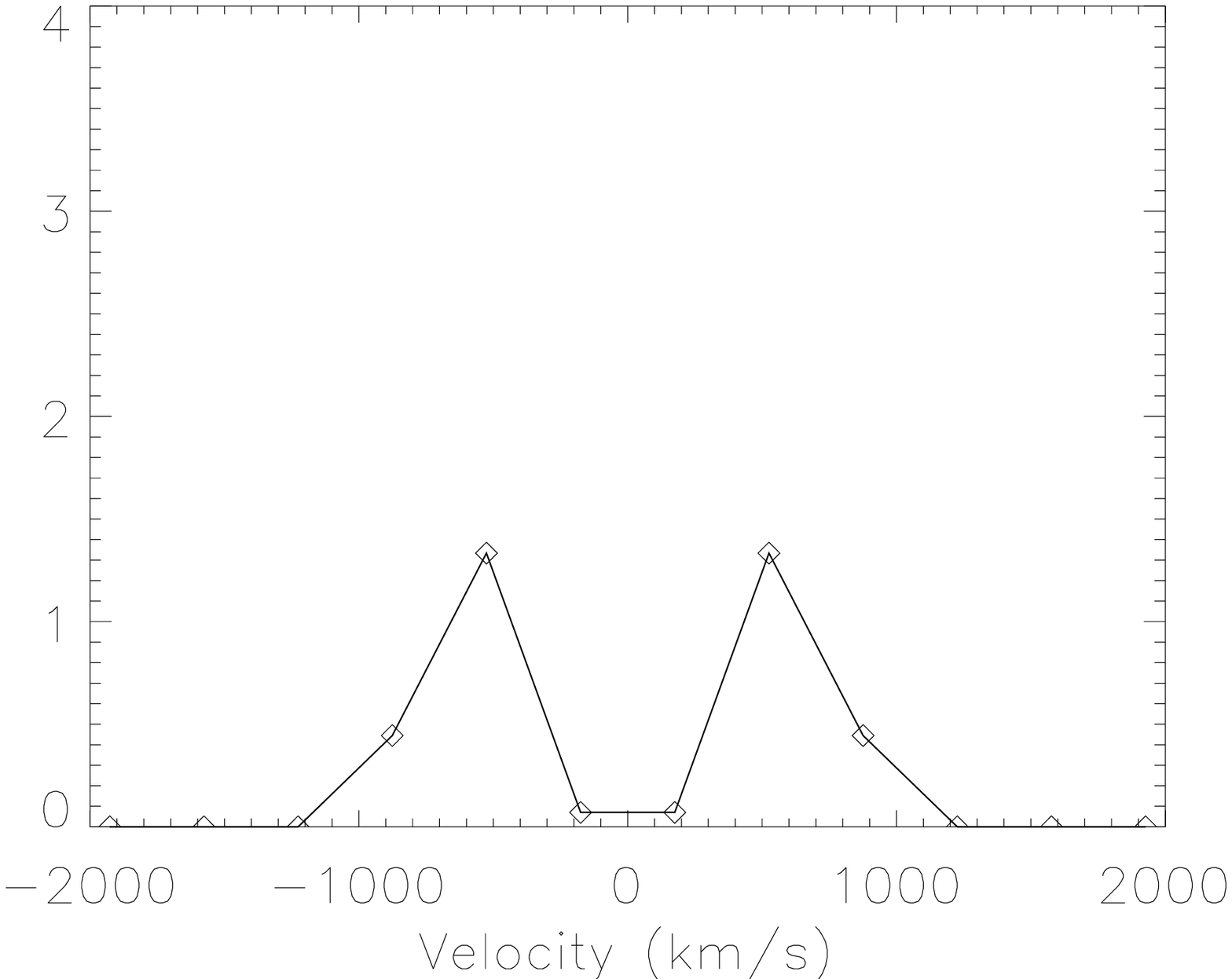}
\FigureFile(40mm,45mm){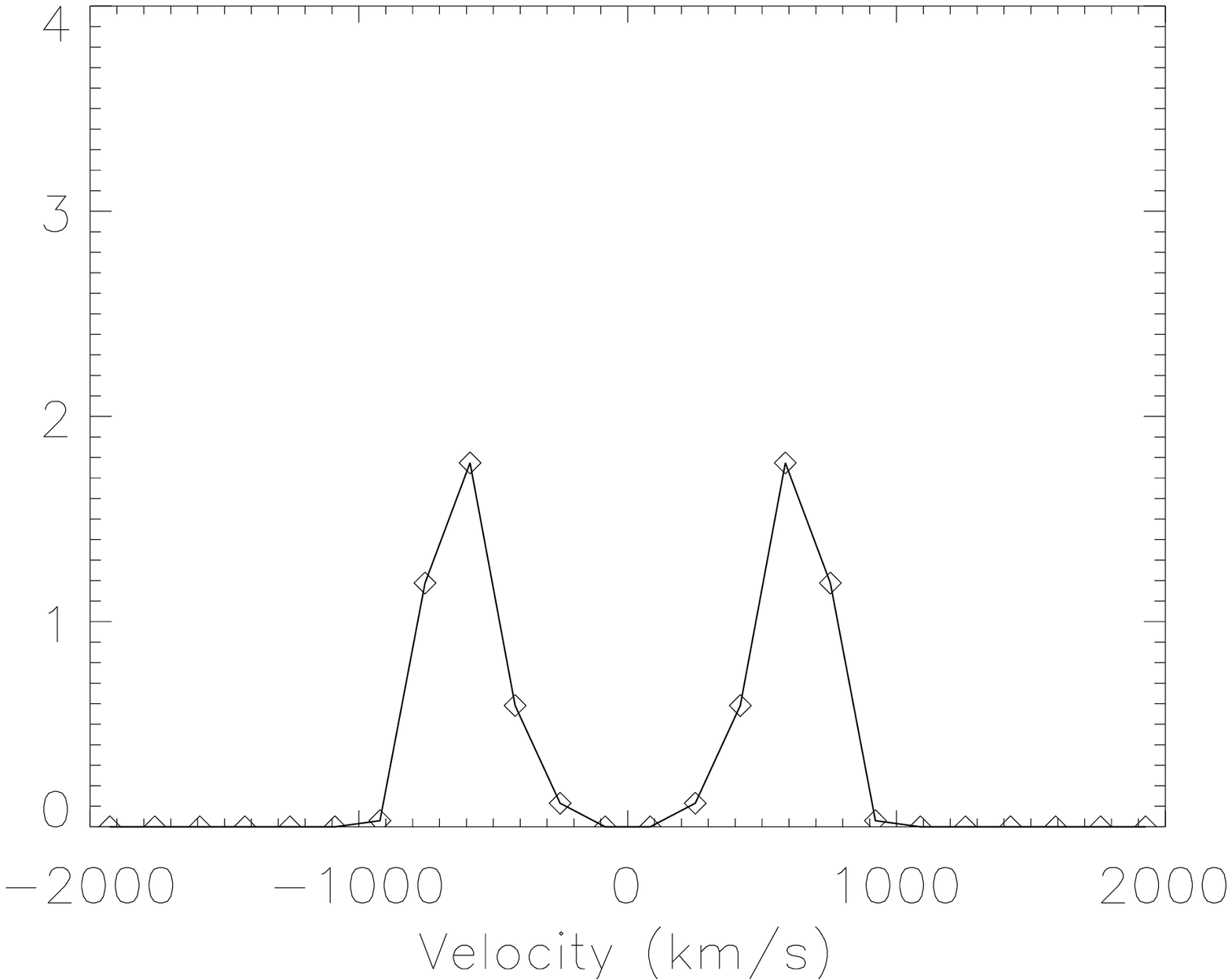}\\
\FigureFile(40mm,35mm){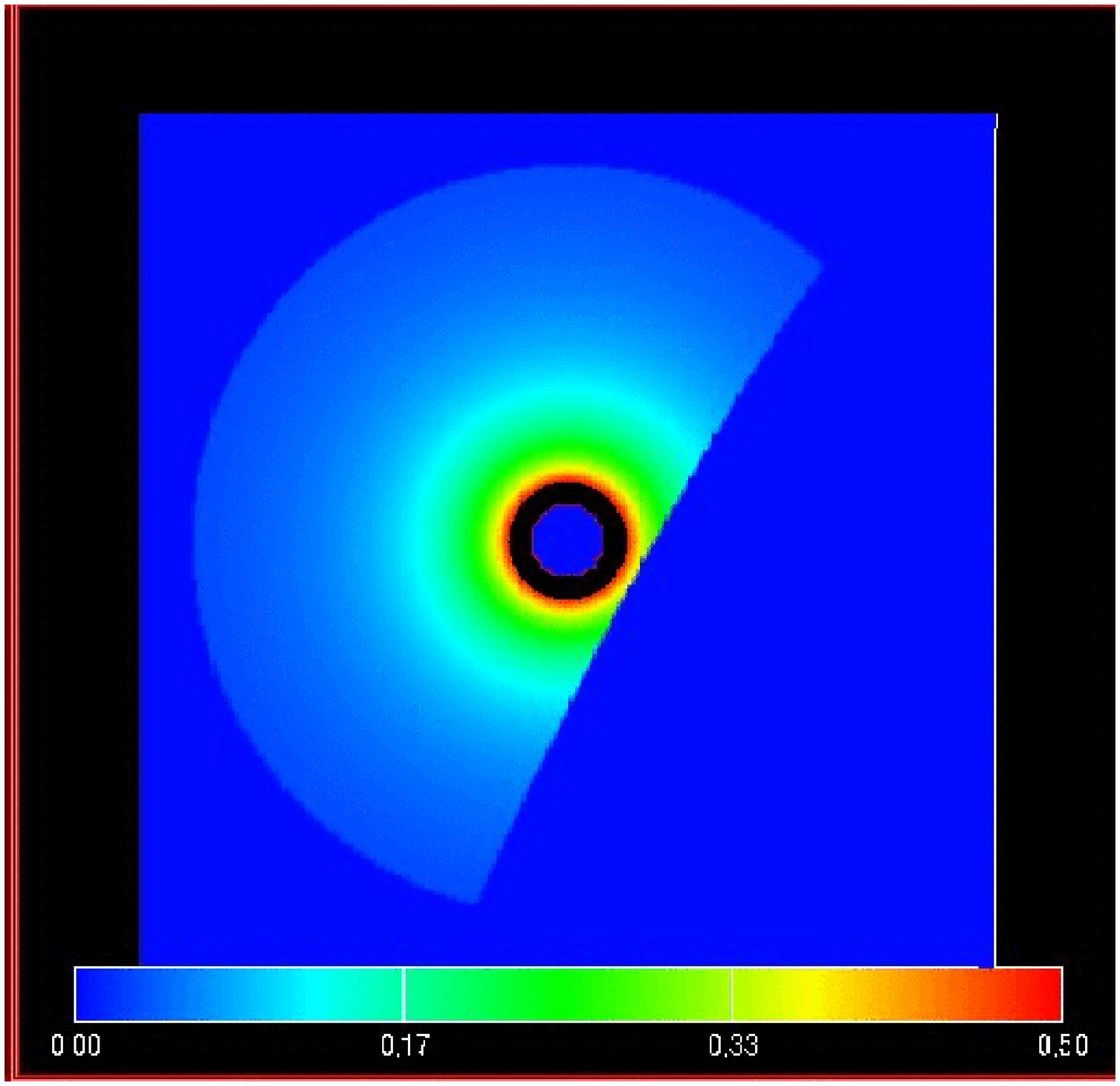}
\FigureFile(40mm,45mm){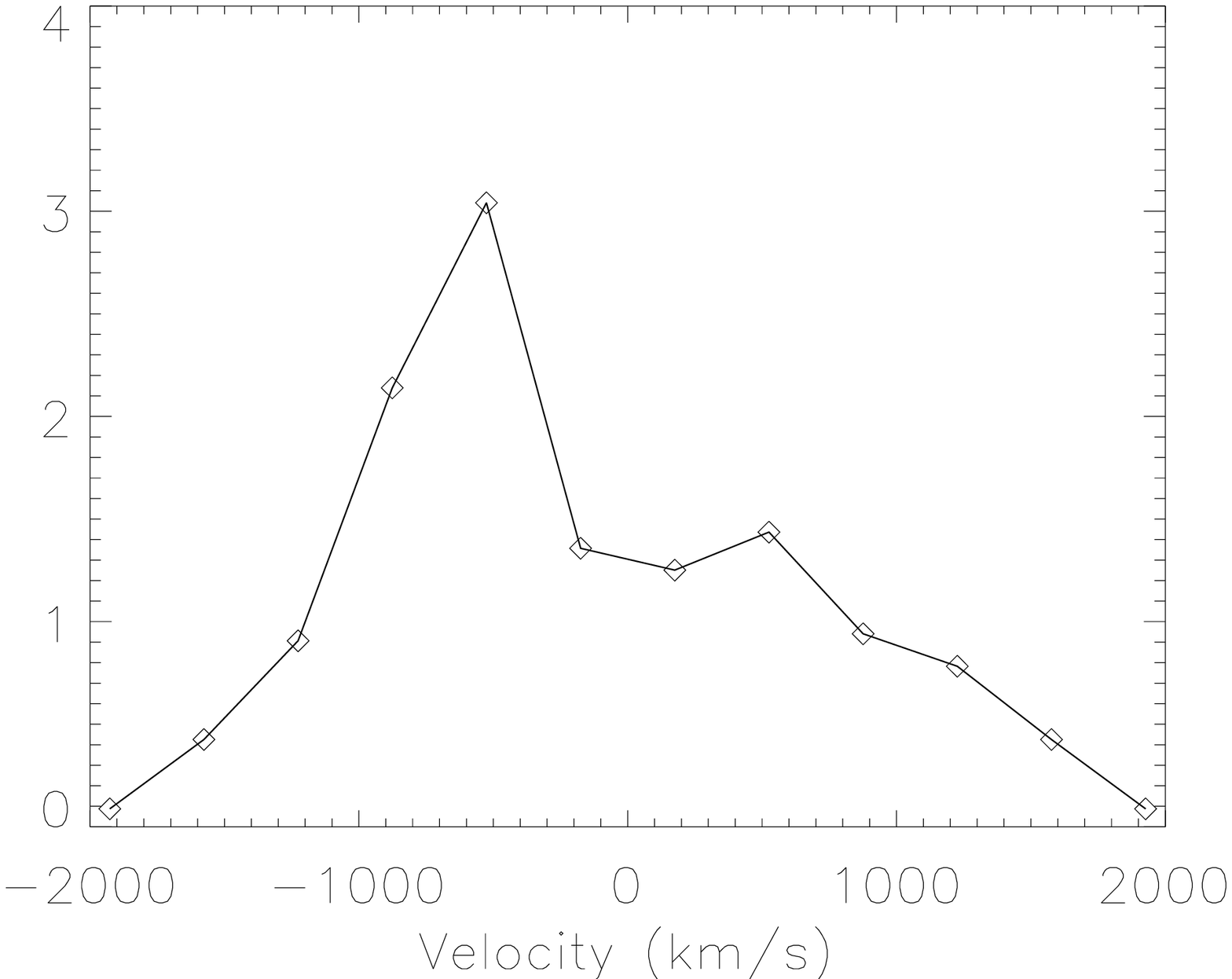}
\FigureFile(40mm,45mm){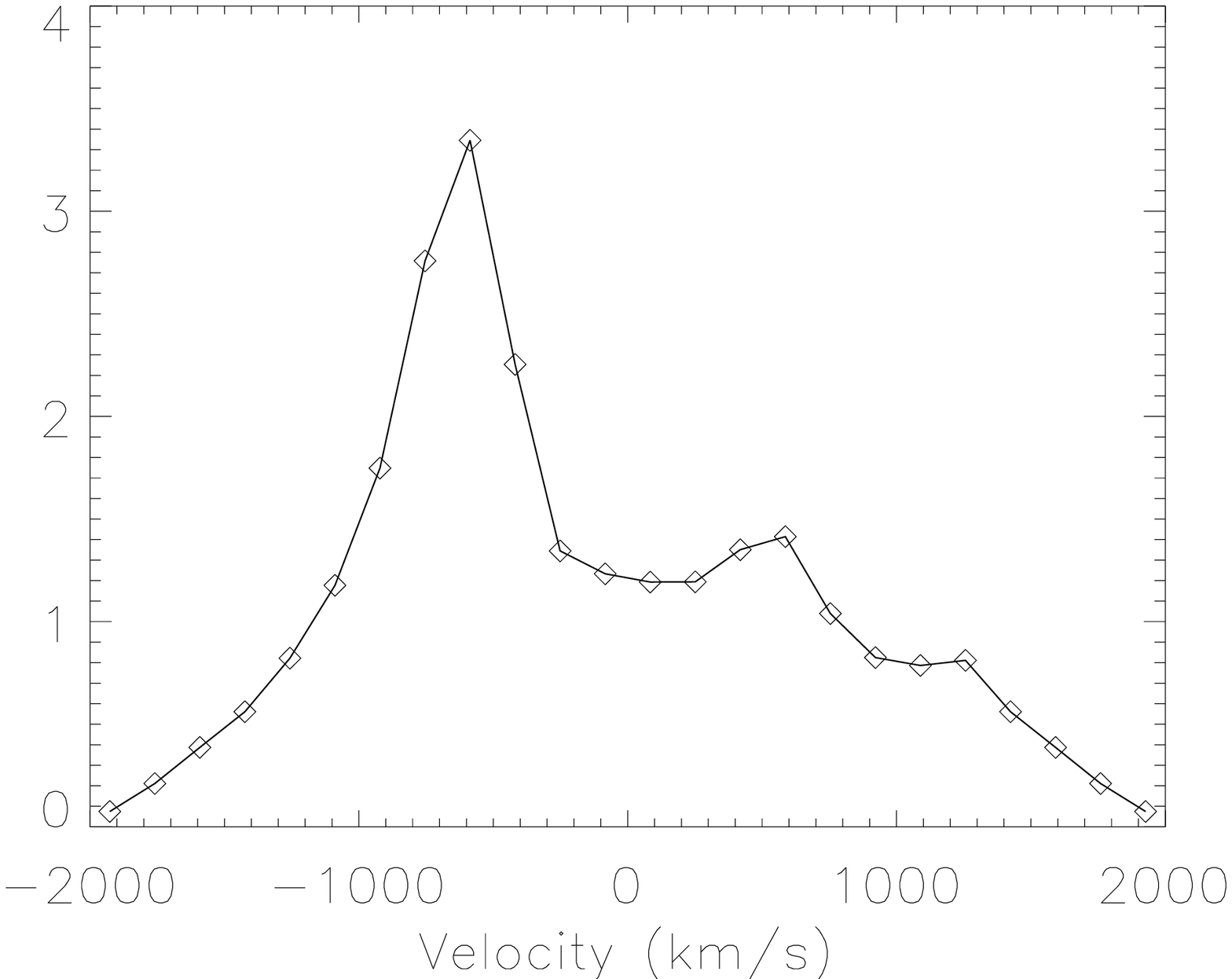}\\
\FigureFile(40mm,35mm){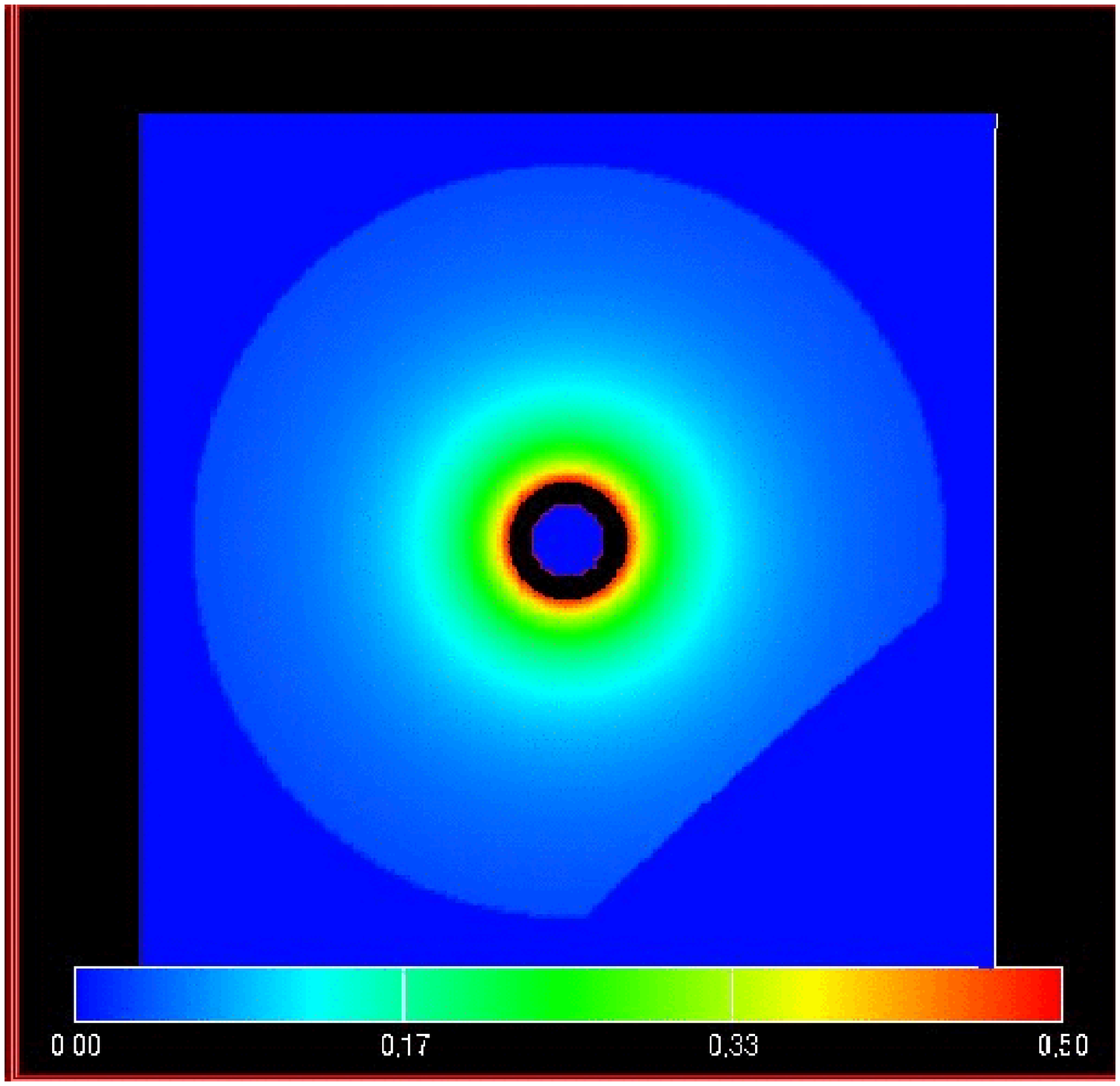}
\FigureFile(40mm,45mm){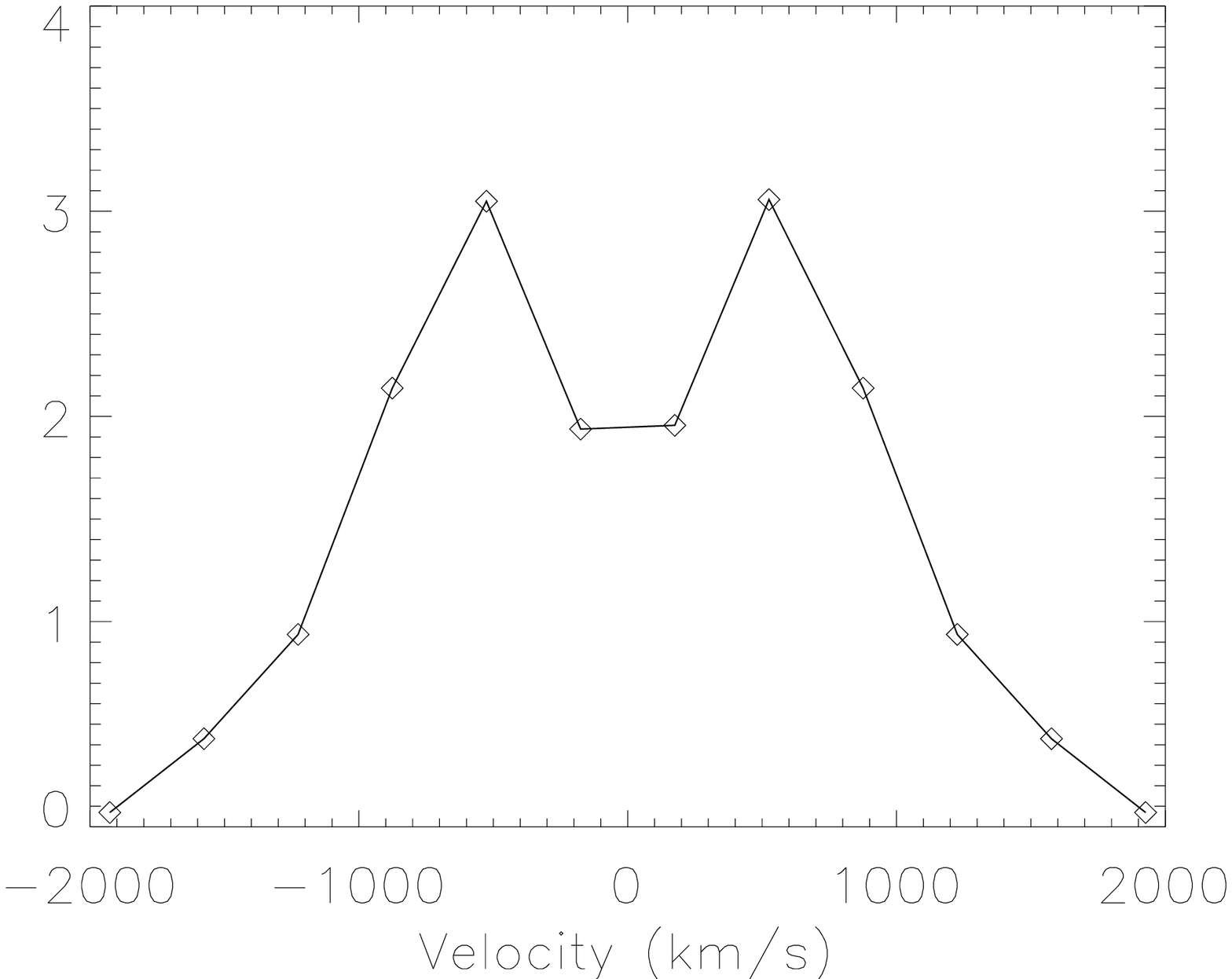}
\FigureFile(40mm,45mm){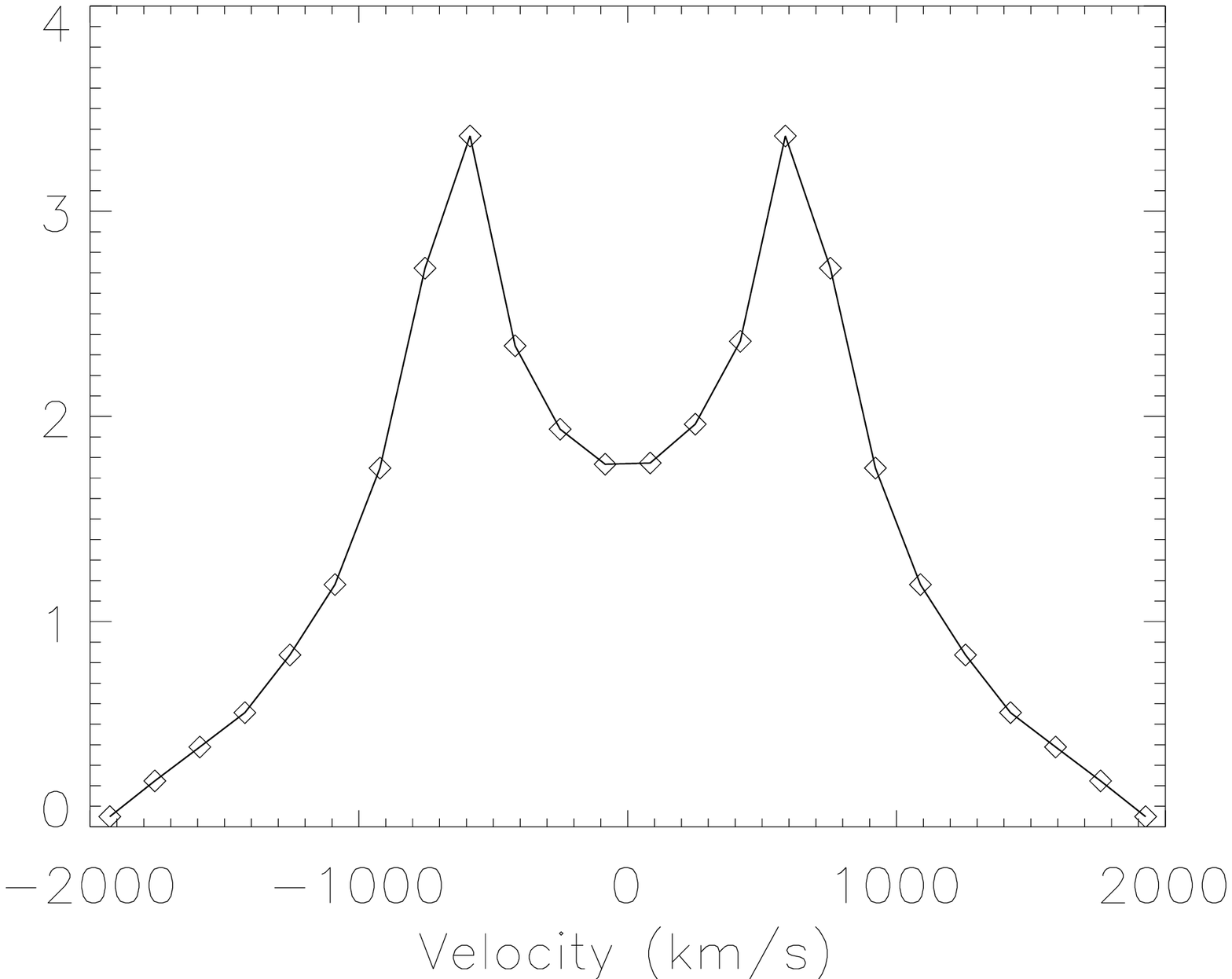}
\end{center}
\caption{Eclipsed disk images (left), emission line profiles
[$F_{\rm line}(\lambda$)]
 for the case of 12 wavelength bins across the line profile
 (i.e. $N=$12; middle), 
and those of $N=$24 (right), respectively,
 for five different binary phases; from the top to bottom,
 $\phi=-0.12, -0.06, 0.00, 0.06$ and 0.12, respectively. 
 The disk is seen by a viewer at the bottom of the page.}
\label{Figeclphase}
\end{figure*}

We first calculated the expected light curves 
based on the line-profile changes
 for the simple cases of $N=12$ and 24; 
i.e., we supposed that 12  (or 24) light curves at 
12 (24) different wavelengths are available.
Figure \ref{Figeclphase} show the images of the eclipsed disk (left)
and the calculated emission line profiles for 
the cases of $N=12$ (middle) and for those of $N=24$ (right), respectively,
for five different binary phases; from the top, 
$\phi=-0.12, -0.06, 0.0, 0.06$, and 0.12, respectively.
Here,
the binary phase of $\phi=0$ corresponds to the mid-eclipse phase.
Note that the binary rotates counterclockwise. 

There are several features which deserve attention.
First,
most of the line profiles exhibit double peaks, and
the wavelengths of the two peaks do not vary significantly
(see, e.g., those in the first, third, and fifth rows).
In contrast, the line wings are mostly missing 
in the third row because of the eclipse of
the innermost part of the disk, which has large rotation velocities.
Second,
the line profiles before and after the eclipse of the central region
(i.e., in the second and fourth rows)
still show two peaks at the same wavelengths as those in the other panels.
In addition, one more weak peak is found in the right panel 
with finer wavelength bins ($N=24$).
Such subtle spectral-line changes are important for 
successful emission-line mapping,
thus a huge number of photons are required.
Obviously, double-horned line profiles are smoother 
in $N=24$ than those in $N=12$.
Finally,
line profiles at one phase (say, $\phi = \phi_0$)
and those at another phase with opposite sign ($\phi = -\phi_0$) 
are exactly symmetric with respect to the line center.
Thus, the light curves at one wavelength (say, at $\lambda_0 + \Delta\lambda$)
are exactly symmetric in time with respect to the minimum
with those at another wavelength at $\lambda_0 - \Delta\lambda$.
This may not be the case in realistic situations
because of the finite thickness effects and possible warps in the disk.

\begin{figure*}
\begin{center}
\FigureFile(38mm,35mm){fig3-1l.eps}
\FigureFile(38mm,45mm){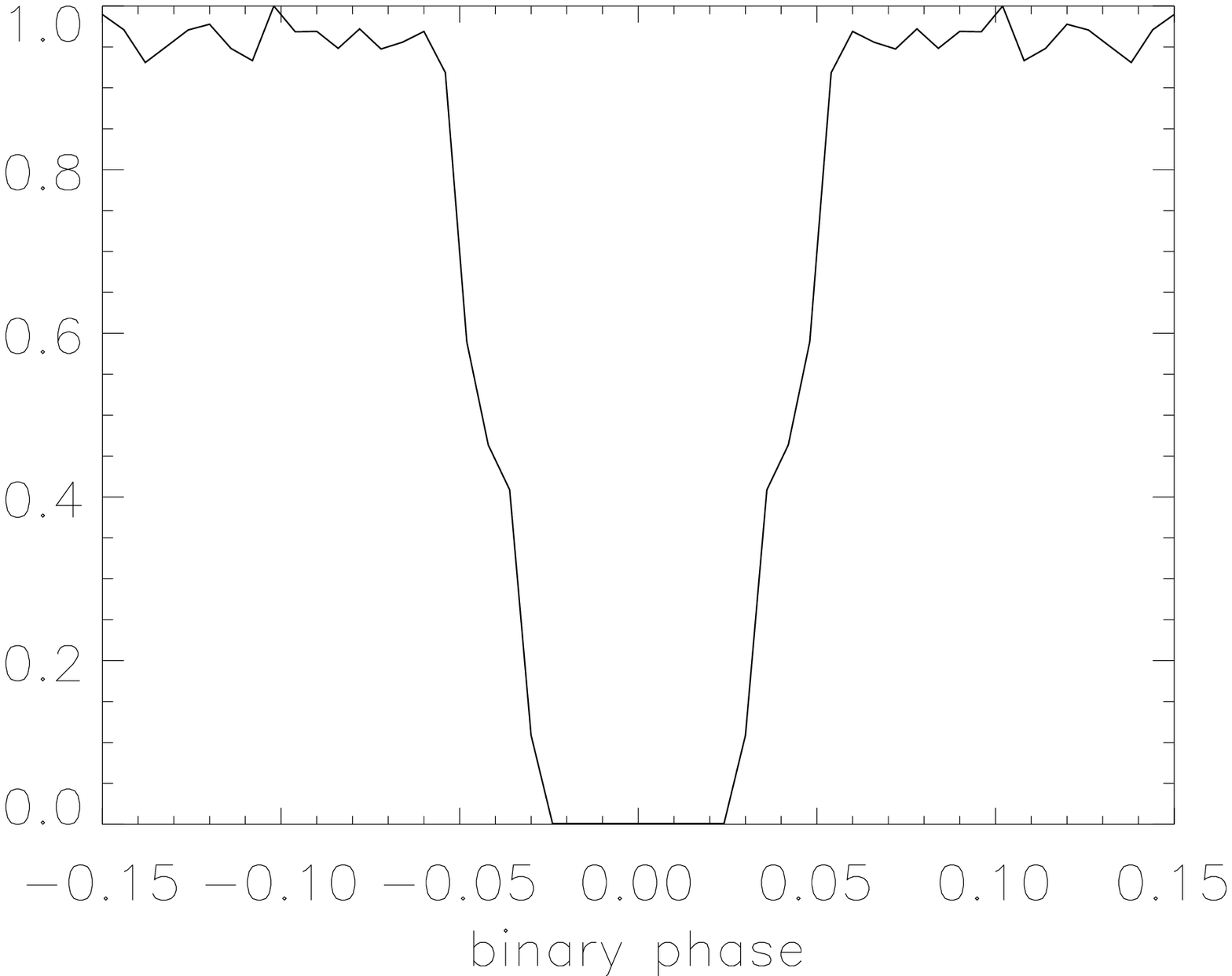}
\FigureFile(38mm,35mm){fig3-1r.eps}\\
\FigureFile(38mm,35mm){fig3-2l.eps}
\FigureFile(38mm,45mm){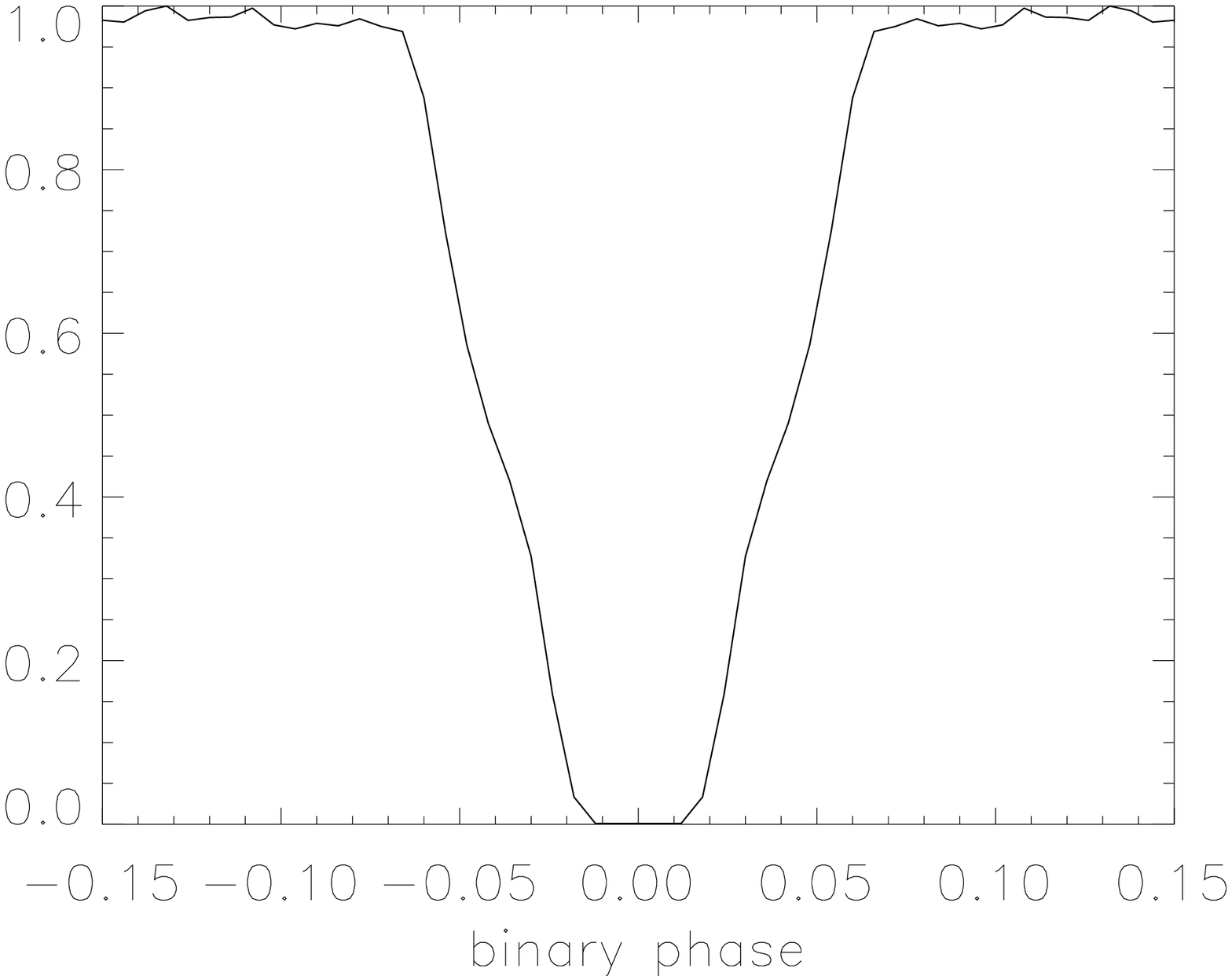}
\FigureFile(38mm,35mm){fig3-2r.eps} \\
\FigureFile(38mm,35mm){fig3-3l.eps}
\FigureFile(38mm,45mm){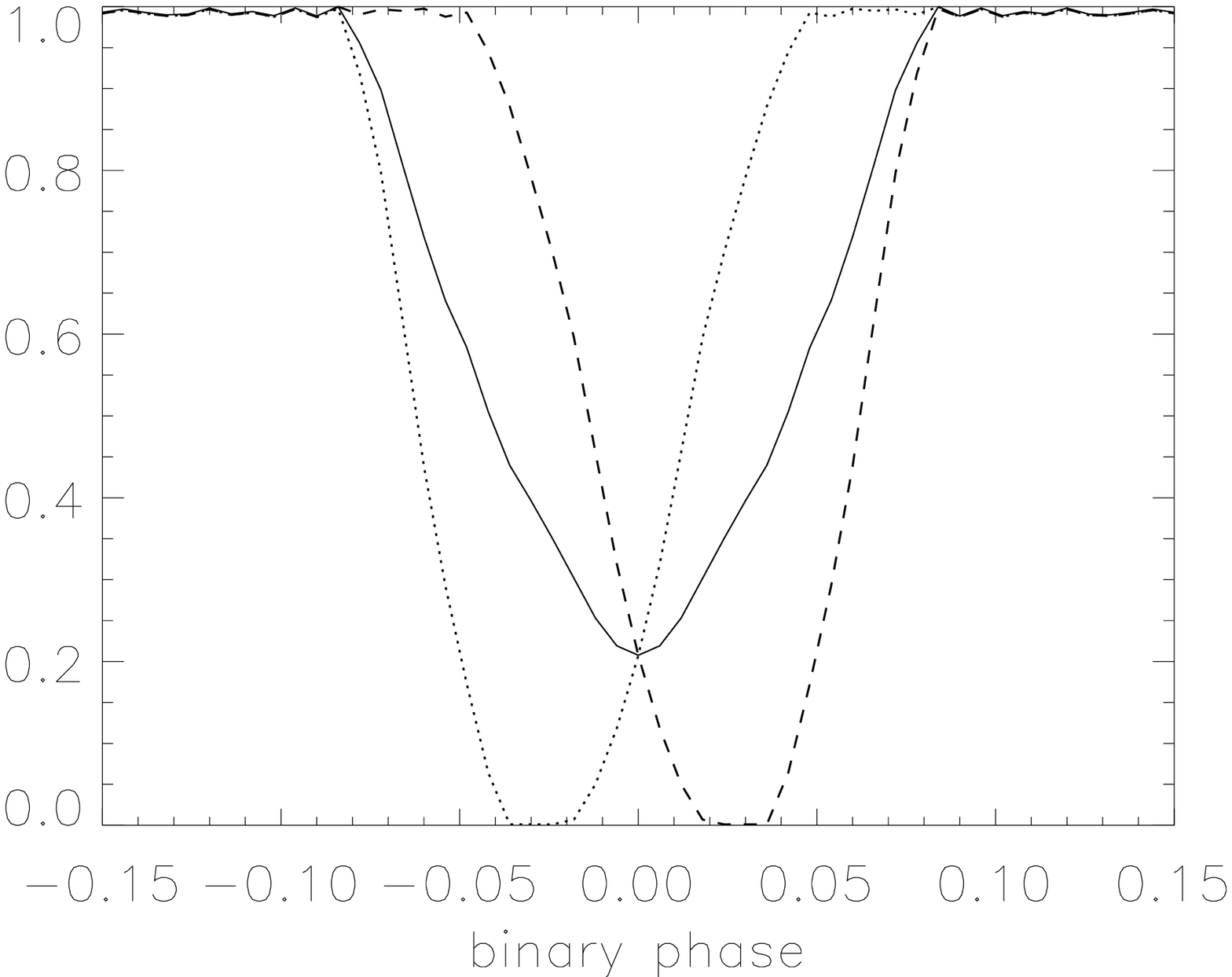}
\FigureFile(38mm,35mm){fig3-3c2.eps}
\FigureFile(38mm,35mm){fig3-3r.eps}\\
\FigureFile(38mm,35mm){fig3-4l.eps}
\FigureFile(38mm,45mm){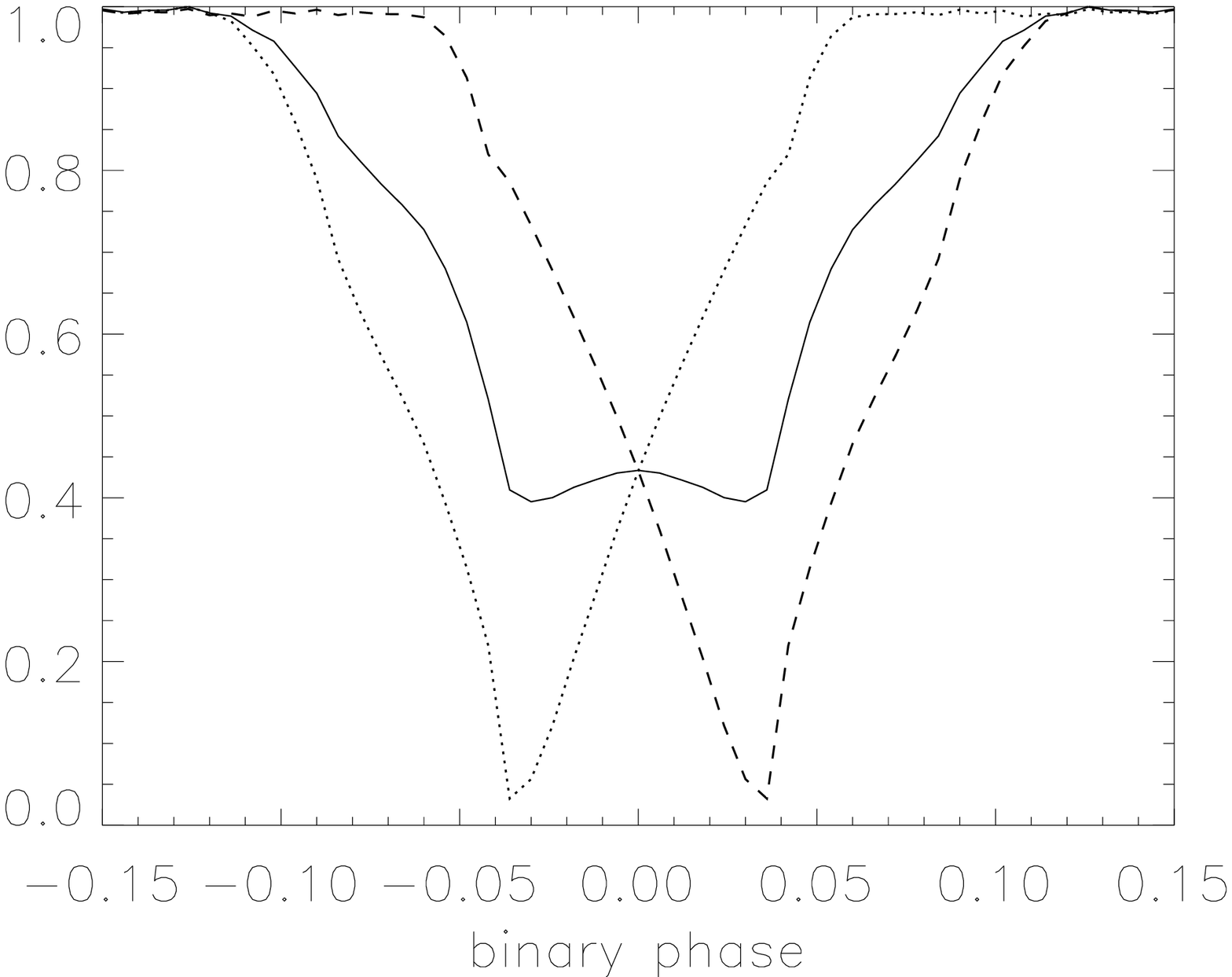}
\FigureFile(38mm,35mm){fig3-4c2.eps}
\FigureFile(38mm,35mm){fig3-4r.eps}\\
\FigureFile(38mm,35mm){fig3-5l.eps}
\FigureFile(38mm,45mm){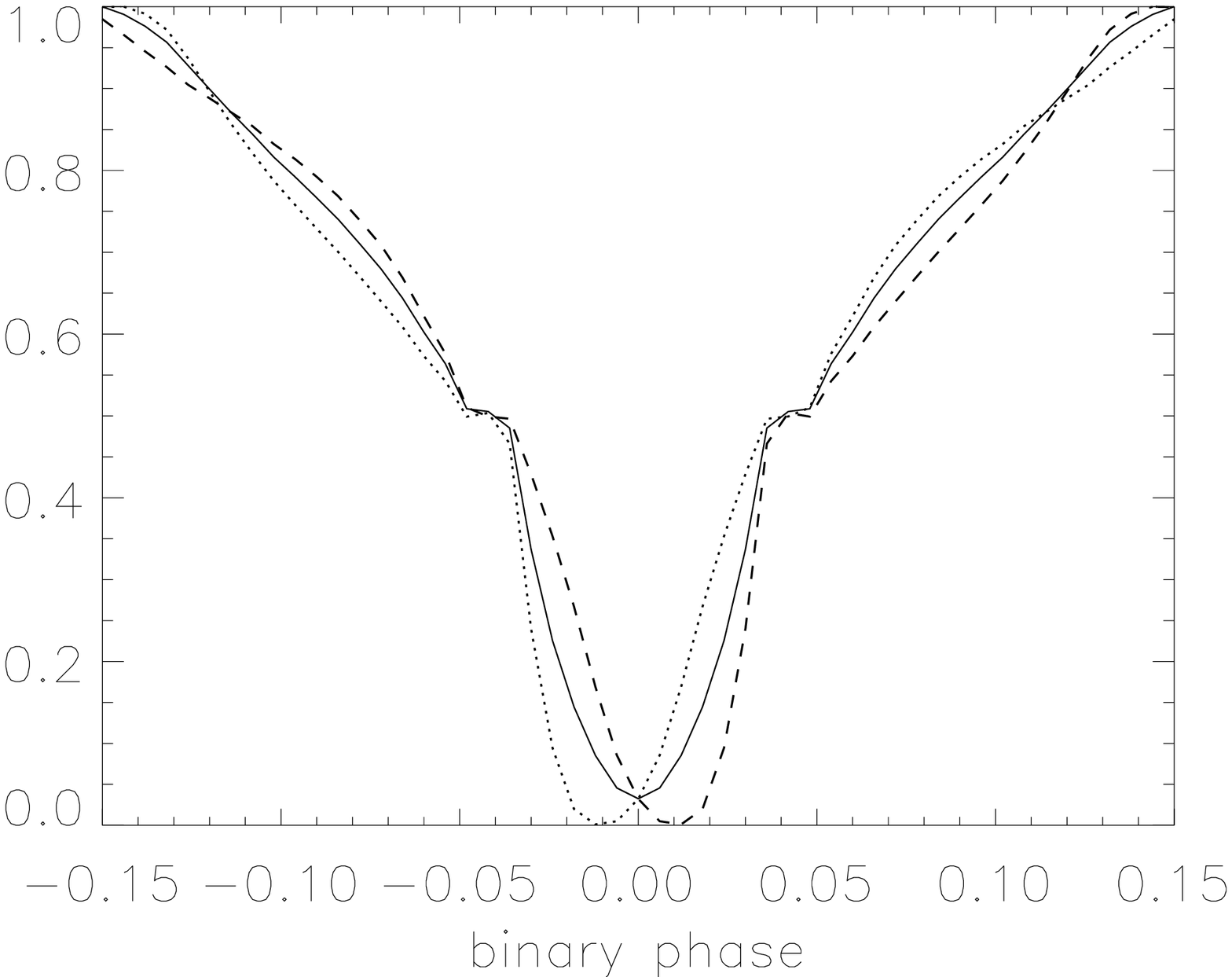}
\FigureFile(38mm,35mm){fig3-5c2.eps}
\FigureFile(38mm,35mm){fig3-5r.eps}\\
\end{center}
\caption{
Original emissivity maps (left panels),
synthetic light curves (middle left), and
the reconstructed maps (middle right) for the cases of $N=12$. 
The right panels in the lower three rows are
reconstructed maps derived solely from the red side of the line profiles.
Each row of the panels represents, from the top to bottom, 
the line-of-sight velocities of $v_{\rm l.o.s.} =$
1400--1750, 1050--1400, 700--1050, 350--700, and 0--350 km s$^{-1}$, 
respectively.
In the light curves, the solid, dotted and dashed lines show those of the sum of the blue and red sides, blue, and red side, respectively.
The disk is seen by a viewer at the bottom of the page.
}
\label{Figlem1}
\end{figure*}

We next display in figure \ref{Figlem1}  
the original maps (left),
synthetic light curves (middle left), and
the reconstructed maps obtained by using both of the red and blue sides
of the line profiles (middle right), 
and those reconstructed by using only the red side (right), 
respectively.
We used $N=12$ wavelength bins for these mappings.
Since the light variations of the red and blue sides are symmetric, 
as mentioned above,
we simply summed up the two contributions in the light curves
to acquire good photon statistics.
The middle panels show a diversity of light curves at different wavelengths.  
The upper two panels representing the cases of high velocities
show similar variations to those of the continuum,
but with a shorter duration, since they have no
complicated shape of the emission region.
As we look downward, the corresponding velocities decrease
and the images start to show a complex pattern.
Accordingly, the light curves are changed in a complex fashion.
Especially, the light curves at the minimum
is nearly flat in the upper three rows
(although the duration of the minimum decreases downward),
which contrasts rather complex variations in other two rows.
This feature suggests that the high-velocity parts of
the line-profile variations seem to be easier to derive
any concrete information.

Eclipse maps are constructed with $21\times 21$ grids.
The constructed maps displayed in the middle right panels
of the upper two rows
show the characteristic {\lq}two-eye' patterns,
which recover the qualitative features of the original maps.
In the lower three rows representing the cases of low line-of-sight velocities,
in contrast, the constructed images (depicted in the middle right panels)
hardly recover the original map. This is because
we used an axisymmetric default image in the reconstruction, whereas
the line emission profile of the original map is far from being axisymmetric.

To fix such a problem, we only use the red side of the line profile
and show the results in the right panels of the lower three rows 
in figure \ref{Figlem1}.  Interestingly, 
the right panels can recover a part of the eclipsed region 
on one side more clearly.
In the third row, especially,
the right panel shows a clear eye-shaped pattern,
which is not seen in the middle right one.
This, therefore, suggests that we are able to obtain the information of
velocity for the case of $v_{\rm l.o.s,}=$ 700--1050 km s$^{-1}$ even
in $N=12$.

\subsection{Line Emissivity Distributions}

\begin{figure*} 
\begin{center}
\FigureFile(75mm,75mm){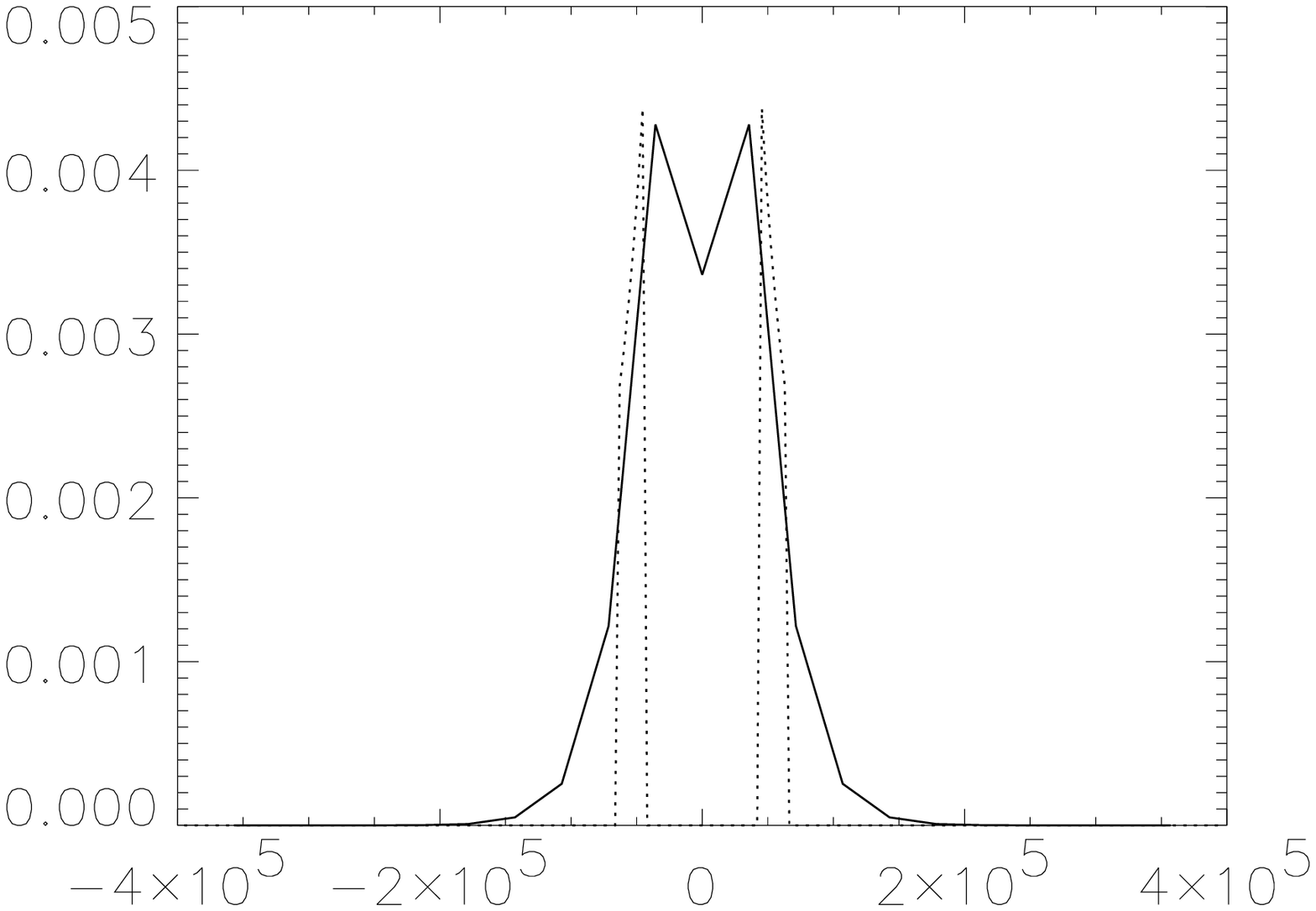}
\FigureFile(75mm,75mm){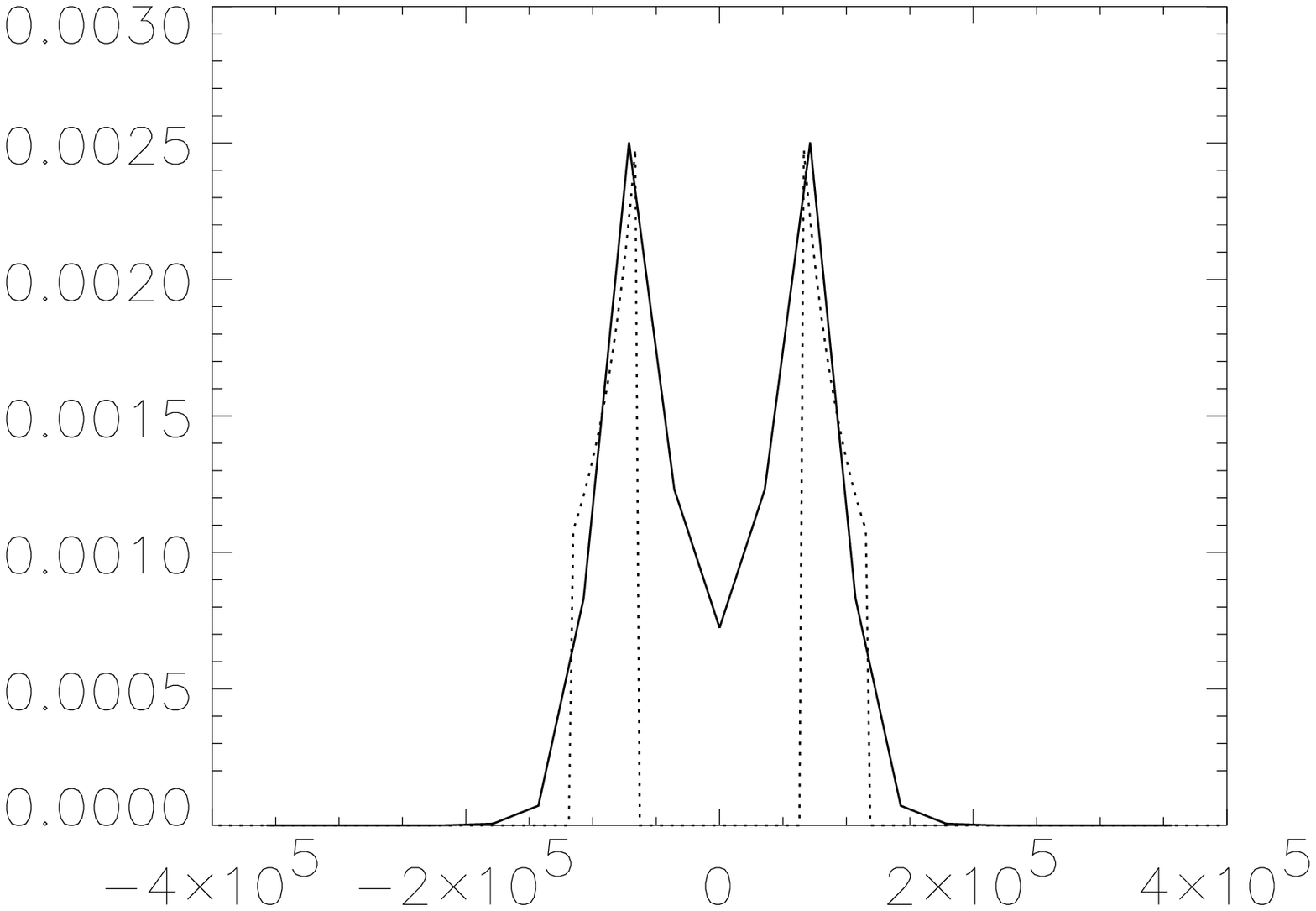}
\end{center}
\caption{One-dimensional distributions of the original (dotted)
and the reconstructed (solid) line emissivity profiles along the $x$-axis
for $v_{\rm l.o.s.} = $ 1400--1750 km s$^{-1}$ (left panels) 
and 1050--1400 km s$^{-1}$(right panels), respectively.
The flux of the eclipse maps is multiplied by 25.}
\label{Figmedam}
\end{figure*}

To see line-emissivity distributions more quantitatively,
we illustrate in figure \ref{Figmedam}
cross-sectional views of the emissivity profile along the $x$-axis.
The emissivity profiles of the original map are illustrated by
the dotted lines, while those of the reconstructed map
are by the solid lines.
Although the reconstructed one shows a broader distribution,
the peak wavelengths do not differ much in both panels.
This figure confirms that
the spatial positions of the two {\lq}eyes' in the reconstructed images
are to good accuracy equal to those of the original ones.
That is, the reconstruction is quite successful at least
as long as data of $\sim 50$ observational runs over one eclipse
are available with good photon statistics.

\subsection{Models with Large Number of Grids}
\begin{figure*}
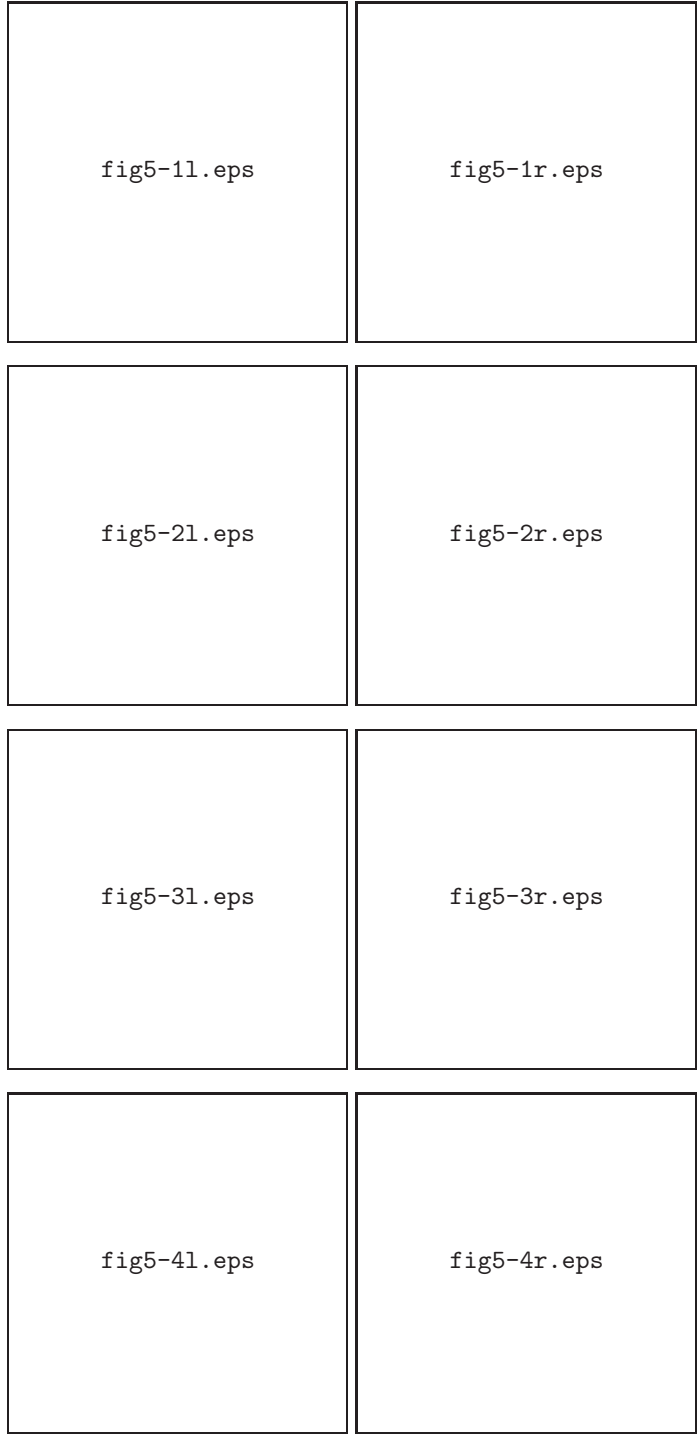

\begin{center}
\FigureFile(45mm,45mm){fig5-1l.eps}
\FigureFile(45mm,45mm){fig5-1r.eps}\\
\FigureFile(45mm,45mm){fig5-2l.eps}
\FigureFile(45mm,45mm){fig5-2r.eps}\\
\FigureFile(45mm,45mm){fig5-3l.eps}
\FigureFile(45mm,45mm){fig5-3r.eps}\\
\FigureFile(45mm,45mm){fig5-4l.eps}
\FigureFile(45mm,45mm){fig5-4r.eps}\\
\end{center}
\caption{
Reconstructed maps for the case of $N=24$,
left panels show models
with $21\times 21$ grids, while the right ones show
those with $41\times 41$ grids.
The line-of-sight velocities are, from top to bottom, 
$v_{\rm l.o.s.}=$ 1750--1925, 1400--1575, 1225--1400, and 875--1050 km s$^{-1}$, respectively.
The disk is seen by a viewer at the bottom of the page.}
\label{Fig24mem}
\end{figure*}

In figure \ref{Fig24mem}, we present only the eclipse maps which show the
characteristic {\lq}two-eye' pattern for the case of $N=24$.
The left panels are for $21\times 21$ grids, as in
figure \ref{Figlem1}, 
while the right panels show finer maps with $41\times 41$ grids. 
The data for $v_{\rm l.o.s.}=$ 1575--1750 km s$^{-1}$
 are not shown, since no convergence is reached in the latter case.
Comparing the left panels with the right ones, we notice that
finer maps represent more circular images.   
However, the positions of two `eyes' do not differ significantly.

\subsection{The Rotation Law}

\begin{figure}
\begin{center}
\FigureFile(85mm,85mm){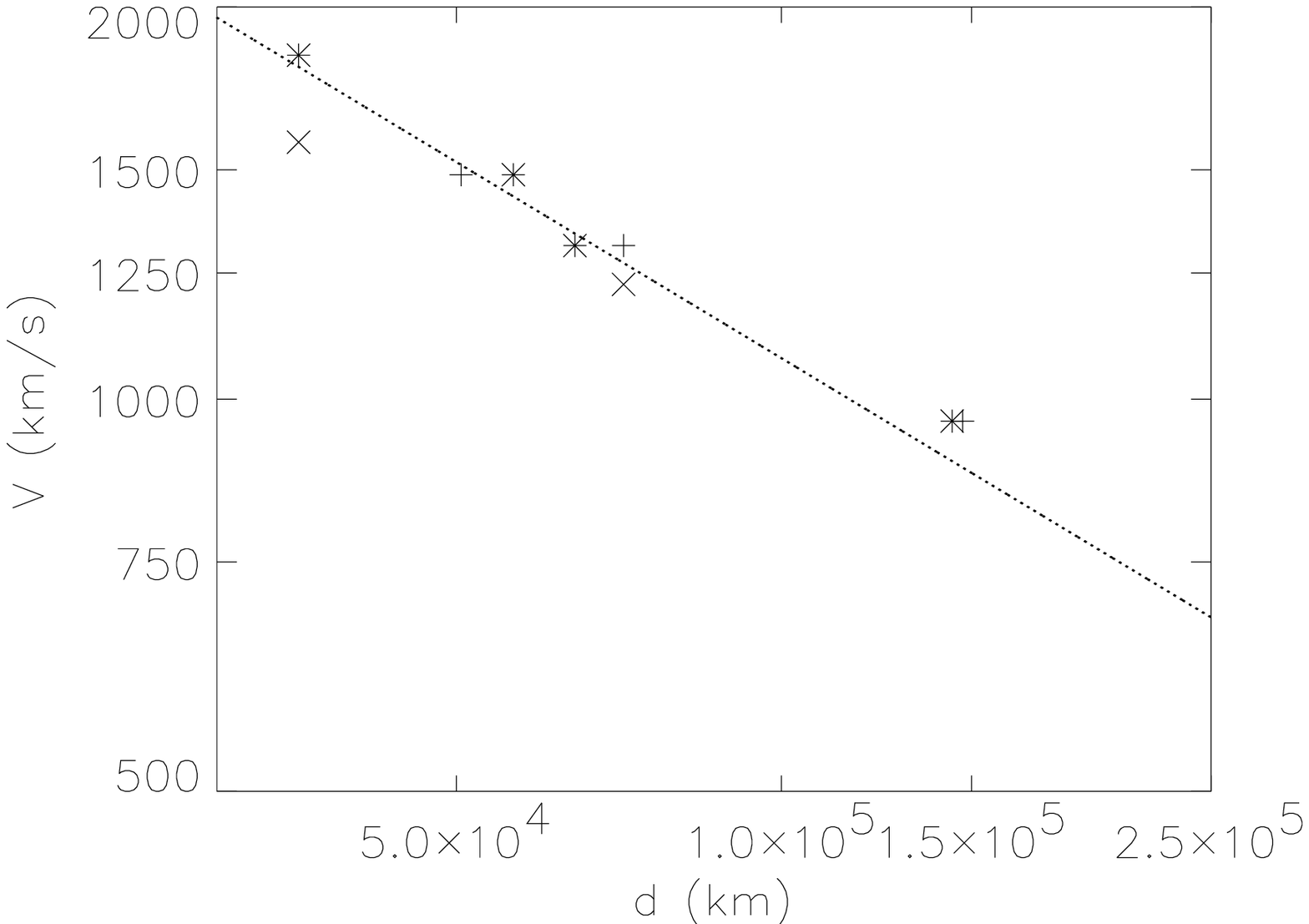}
\end{center}
\caption{
Line-of-sight velocities obtained by the image reconstruction technique
as functions of the distance from the primary star. 
The results for the case of $N=12$ are indicated by $\lq{+}$',
while those of $N=24$ are by $\lq\times${\rq}, respectively,
both obtained with $21 \times 21$ grids. 
$\lq\ast${\rq} also shows the  results for the case of $N=24$ and $41 \times 41$ grids.  
For a comparison, we depict the Keplerian velocity relation,
$v\propto r^{-1/2}$, by the dotted line.
}
\label{Figkep}
\end{figure}

Finally, we calculated, for each line-of-sight velocity,
the distance from the primary star in the real space
based on the separation between the two `eyes' on the reconstructed image;
the result is plotted in figure \ref{Figkep}.
As is clear, the reconstructed line-of-sight velocities are roughly
on the straight, dotted line with a slope in proportion to $d^{-1/2}$.
That is, our proposed method is feasible to reconstruct Keplerian velocity fields, as long as good photon statistics is achieved.

To see how observational errors affect our results, 
we repeated the same procedure,
but added random errors in calculating the light curves at each wavelength. 
Namely, we calculate the observable line flux, $F_{\rm line}^{\rm obs}$, 
from the theoretically modeled flux, $F_{\rm line}^{\rm model}$
according to the following steps:\\
\begin{enumerate}
\item $F_{\rm tot}^{\rm model}(\lambda) 
                \equiv F_{\rm line}^{\rm model}(\lambda) + F_{\rm cont}$,
\item $F_{\rm tot}^{\rm obs}(\lambda) = F_{\rm tot}^{\rm model}(\lambda) 
         + \epsilon({\rm S/N})^{-1}\sqrt{F_{\rm tot}^{\rm model}(\lambda)}$,
\item $F_{\rm line}^{\rm obs}(\lambda) 
                \equiv F_{\rm tot}^{\rm obs}(\lambda) - F_{\rm cont}$.
\end{enumerate}
Here, $\epsilon$ represents the Gaussian noise (with the mean being zero
and the standard deviation being unity); the amount of random variations
is given by the signal-to-noise ratio (S/N).
Note that we set $F_{\rm cont} = 1.0$, and $F_{\rm line}^{\rm model}$ was
calculated so as to give an equivalent width of 100\AA 
(see subsection 2.3).

\begin{figure*}
\begin{center}
\FigureFile(70mm,70mm){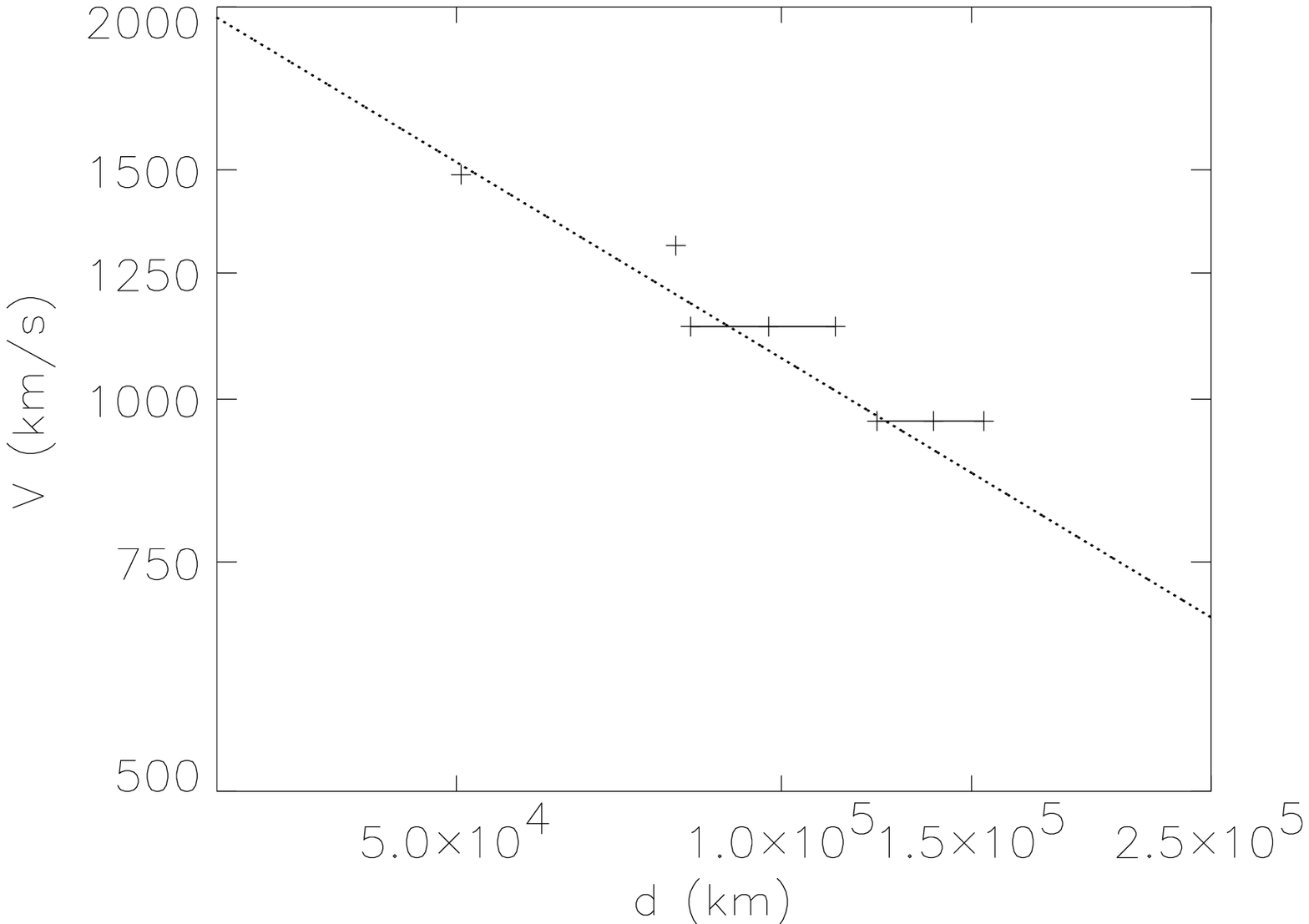}
\FigureFile(70mm,70mm){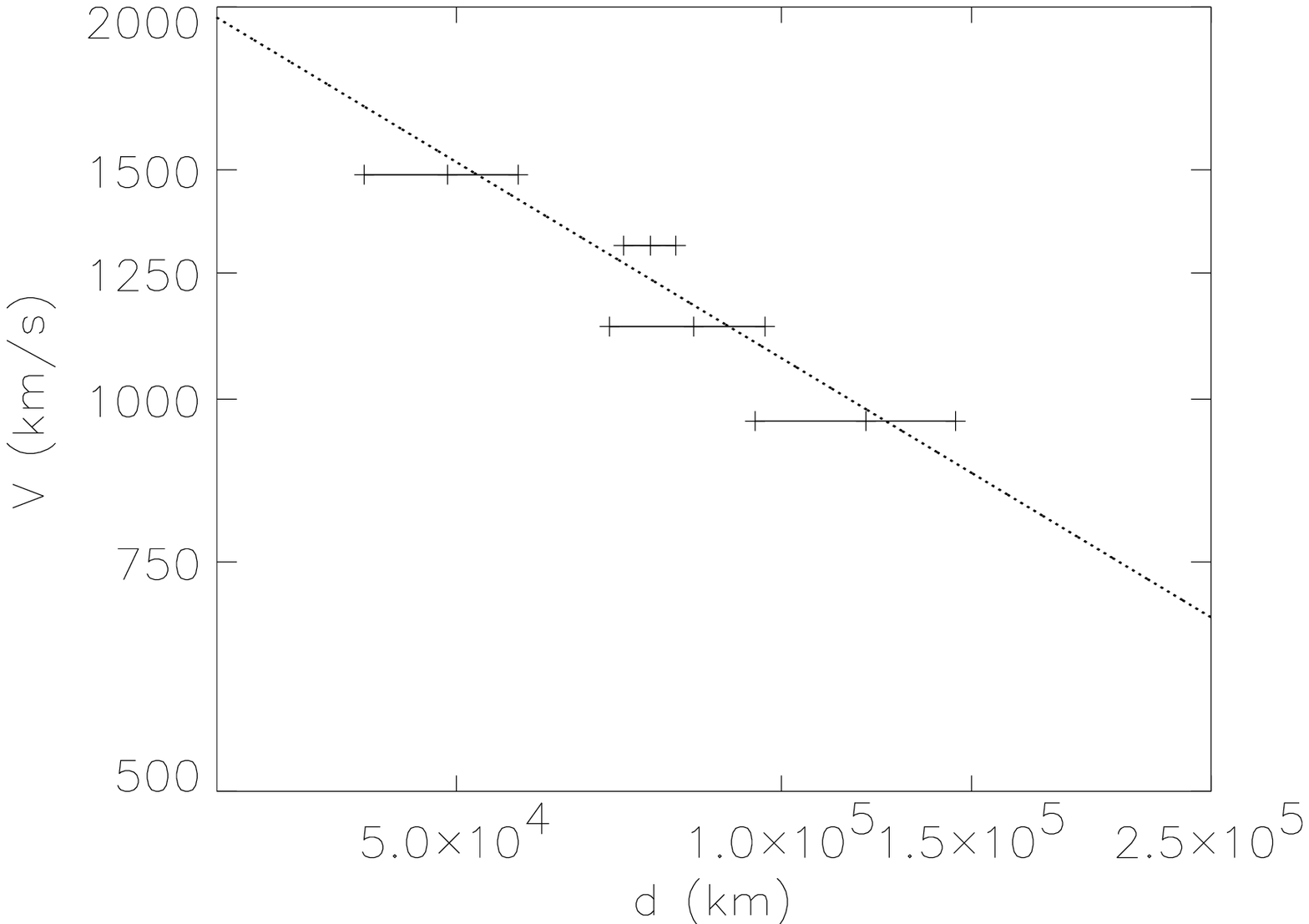}
\end{center}
\caption{Effect of noise in the case of $N({\rm wavelength ~bins})=24$. 
In the case of S/N = 10, 
the left panel shows the result with $21 \times 21$ grids for 
the reconstruction and the right panel does the one with $41 \times 41$ grids.  
For a comparison, we also depict the Keplerian velocity relation,
$v\propto r^{-1/2}$, with the dotted line.
}
\label{Fignoi}
\end{figure*}

We calculated the cases of S/N = 10, 20, and 50 as representative examples.
We found that in some cases convergence is not always guaranteed 
in the eclipse mapping process, because the light curve is not symmetric
when random errors are added.
Nevertheless, we expected good results, even if we solely consider the
cases with successful reconstruction.
Figure \ref{Fignoi} presents the results of the relatively noisy cases 
of S/N = 10.  The number of wavelength bins is $N= 24$.  
It is impressive to see that
the Keplerian relation is roughly reproduced,
although we used only those results which have the clear {\lq}eye' pattern.
These results demostrate that our proposed method is powerful even for 
the case of S/N = 10, as long as the reconstructed eclipse map
displays the {\lq}eye' pattern.

\section{Discussion}
We propose {\it emission-line eclipse mapping} to map the velocity
fields in an accretion disk. We studied the feasibility of emission-line 
eclipse mapping using model simulations. 
Our results show that those regions having high line-of-sight velocities
can be precisely reconstructed, while the low line-of-sight regions
are not, as long as an axisymmetric default image is used.
These reconstructed maps of high line-of-sight velocities show
the characteristic {\lq}two-eye' pattern.
The corresponding line-of-sight velocity recovers the
originally adopted relation, which is proportional to $d^{-0.5}$.
It means that this method is a potentially useful tool to 
probe the Keplerian disk rotation 
from observations of the line-profile variations.

Although the Keplerian rotation is a fundamental assumption
in constructing the so-called standard disk model, it
is not always trivial from a theoretical point of view.
For example,
a rather hot accretion flow, called 
ADAF (advection-dominated accretion flow), predicts
sub-Keplerian rotation (Kato et al. \yearcite{KFM98});
i.e., $v_\varphi < v_{\rm K}$.
In addition, the velocity fields are affected by the presence 
of large-scale magnetic fields and/or spiral patterns, which will produce
deviations from the exact Keplerian velocity of the order of 
$\delta v/v_{\rm K} \sim v_{\rm A}/v_{\rm K} \sim c_{\rm s}/v_{\rm K} 
\sim H/R \simeq 1/30$ (where
$v_{\rm A}$ and $c_{\rm s}$ are the Alfv$\acute{\rm e}$n velocity and the sound velocity,
 respectively,
and $H/R$ is the aspect ratio of the disk).
To detect such small effects we need to collect an enormous number of photons,
which are unavailable at the present. 
Thus, we should await innovations of the observation technique.

For successful reconstructions, we need rather frequent 
observations over the eclipse periods.
Only 8-m class telescopes, such as Subaru, can meet this requirement,
which is underway.
The next issues are to improve the current method so as to minimize the
total observing times by modifying the default images
and to perform observations studies 
(e.g., Vrielmann et al. \yearcite{V95}).

\bigskip 
We are grateful to Taichi Kato and Raymundo Baptista for useful discussion and helpful comments.
The author (M. M.) also thanks Makoto Uemura for his advice on using PRIDA code. 
This work was supported in part by the Grants-in Aid of the
Ministry of Education, Culture, Sports, Science and Technology
(13640238, SM).

%%%%%%%%%%%%%%%%%%%%%%%%%%%%%%%%%%%%%%%

%%%
% See the manual for the detail.
%%%


\begin{thebibliography}{}
\bibitem[BS(1991)]{BS91}
 Baptista, ~R., \& Steiner, ~J. ~E. 1991, \aap, 249, 284
\bibitem[BS(1993)]{BS93}
 Baptista, ~R., \& Steiner, ~J. ~E. 1993, \aap, 277, 331
\bibitem[H(1985)]{H85}
 Horne, ~K. 1985, \mnras, 213, 129
\bibitem[H(1993)]{H93}
 Horne, ~K. 1993, in Accretion Disks in Compact Stellar Systems,
  ed.\ J.~C. ~Wheeler (Singapore: World Scientific), 117
\bibitem[H(1994)]{H94}
 Horne, ~K. 1994, in Theory of Accretion Disks -- 2.
  ed.\ W.~J. Duschl, J. Frank, F. Meyer, E. Meyer-Hofmeister, \&
		  W. M. Tscharnuter (Dordrecht: Kluwer Academic Publishers), 77
\bibitem[HM(1986)]{HM86}
 Horne, ~K., \& Marsh, ~T. ~R. 1986, \mnras, 218, 761
\bibitem[HS(1991)]{HS91}
 Horne, ~K., \& Saar, ~S. ~H. 1991, \apj, 374, L55
\bibitem[KFM(1998)]{KFM98}
 Kato, ~S., Fukue, ~J., \& Mineshige, ~S. 1998, Black-Hole Accretion
		  Disks (Kyoto: Kyoto Univ. Press)
\bibitem[MH(1988)]{MH88}
 Marsh, ~T. ~R., \& Horne, ~K. 1988, \mnras, 235, 269
\bibitem[SMH(1986a)]{SMH86a}
 Sawada, ~K., Matsuda, ~T., \& Hachisu, ~I. 1986, \mnras, 219, 75
\bibitem[SMH(1986b)]{SMH86b}
 Sawada, ~K., Matsuda, ~T., \& Hachisu, ~I. 1986, \mnras, 221, 679
\bibitem[YMMFUT(1998)]{YMMFUT98}
 Yonehara, ~A., Mineshige, ~S., Manmoto, ~T., Fukue, ~J., Umemura, ~M.,
\& Turner, ~E.~L. 1998  \apj, 501, L41 (Erratum 511, L65) 
\bibitem[V(1995)]{V95}
 Vrielmann, ~S., Horne, ~K., \& Baptista, ~R. 1995, in  Cataclysmic
		  Variables, ed. \ A. ~Bianchini, ~M. ~della ~Valle,  \&
		  M. ~Orio (Dordrecht: Kluwer Academic Publishers), 376
\end{thebibliography}
\end{document}